\documentclass[twoside,10pt,hidelinks]{article}
\usepackage{tabularx} 
\usepackage{graphicx}
\usepackage[graphicx]{realboxes}
\usepackage[a4paper, total={6in, 8in}]{geometry}

\usepackage{morefloats}
\usepackage{amsfonts}
\usepackage{color}
\usepackage{multirow}\usepackage{multirow}

\usepackage{multirow}
\usepackage{mathtools}

\usepackage{latexsym}
\usepackage{amsmath,amssymb,amsthm}
\usepackage{dsfont}
\usepackage{extarrows}
\usepackage{hyperref}
\usepackage{lscape}
\usepackage{units}
\usepackage{amsmath}
\usepackage{natbib}
\usepackage{pdflscape}
\usepackage{afterpage}
\usepackage{adjustbox}
\usepackage{colortbl}

\numberwithin{equation}{section}

\newtheoremstyle{thm}
{9pt}
{9pt}
{\itshape}
{}
{\bfseries}
{.}
{ }
{}
\theoremstyle{thm}

\newtheorem{theorem}{Theorem}[section]

\newtheoremstyle{def}
{9pt}
{9pt}
{}
{}
{\bfseries}
{.}
{ }
{}
\theoremstyle{def}

\newtheorem*{remark*}{Remark}

\usepackage{color,xcolor}
\renewcommand{\footnoterule}{%
 & \kern -3.5pt
 & \hrule width \textwidth height 1pt
 & \kern 3.5pt
}

\makeatletter
\def\blfootnote{\xdef\@thefnmark{}\@footnotetext}
\makeatother

\begin{document}

\title{\bf Testing for the Pareto type I distribution: A comparative study}


\author{L. Ndwandwe, J.S. Allison, L. Santana, I.J.H. Visagie}

\date{}
\maketitle

\begin{abstract}
{
Pareto distributions are widely used models in economics, finance and actuarial sciences. As a result, a number of goodness-of-fit tests have been proposed for these distributions in the literature. We provide an overview of the existing tests for the Pareto distribution, focussing specifically on the Pareto type I distribution. To date, only a single overview paper on goodness-of-fit testing for Pareto distributions has been published. However, the mentioned paper has a much wider scope than is the case for the current paper as it covers multiple types of Pareto distributions. The current paper differs in a number of respects. First, the narrower focus on the Pareto type I distribution allows a larger number of tests to be included. Second, the current paper is concerned with composite hypotheses compared to the simple hypotheses (specifying the parameters of the Pareto distribution in question) considered in the mentioned overview. Third, the sample sizes considered in the two papers differ substantially.

In addition, we consider two different methods of fitting the Pareto Type I distribution; the method of maximum likelihood and a method closely related to moment matching. It is demonstrated that the method of estimation has a profound effect, not only on the powers achieved by the various tests, but also on the way in which numerical critical values are calculated. We show that, when using maximum likelihood, the resulting critical values are shape invariant and can be obtained using a Monte Carlo procedure. This is not the case when moment matching is employed.

The paper includes an extensive Monte Carlo power study.
Based on the results obtained, we recommend the use of a test based on the phi divergence together with maximum likelihood estimation.
}
\end{abstract}


\section{Introduction and motivation}

The Pareto distribution was first introduced by the economist Vilfredo Pareto in 1897 as a model for the distribution of income, see \cite{Par}. Since then the Pareto distribution has been widely used in a variety of fields including economics, finance, actuarial science, and reliability theory, see, e.g., \cite{nofal2017new} as well as \cite{ismail2004simple}. For an in-depth discussion of the Pareto distribution the interested reader is referred to \cite{Arn} where the role of this distribution in the modelling of data is discussed.


The popularity of the Pareto distribution has prompted research into several generalisations of this model. Subsequently, the originally proposed distribution became known as the Pareto type I distribution in order to distinguish this model from the variants known as the Pareto types II, III and IV as well as the so-called generalised Pareto distribution. These distributions, as well as the relationships between them, are described in detail in \cite{Arn}.

Due to the wide range of applications of the various types of Pareto distributions, a number of tests have been developed for the hypothesis that observed data follow a Pareto distribution. This paper provides an overview of the goodness-of-fit tests specifically developed for the Pareto type I distribution available in the literature. Although numerous overview papers are available for goodness-of-fit tests for distributions such as the normal distribution, see, e.g., \cite{bera2016new}, and the exponential distribution, see, e.g., \cite{allison2017apples}, the only overview paper of this kind relating to the Pareto distribution is \cite{chu2019review}. The latter investigates several existing tests for the Pareto types I and II as well as the generalised Pareto distribution. However, due to the wider scope,  \cite{chu2019review} does not review all of the tests available for the Pareto type I distribution; several recently proposed tests are excluded from the comparisons provided. The current paper has a narrower scope and provides an overview of existing tests specifically for the Pareto type I distribution, hereafter simply referred to as the Pareto distribution.

A further distinction between \cite{chu2019review} and the study presented here is that the former considers simple hypotheses in which the parameters of the Pareto distribution are specified beforehand, whereas the current paper is concerned with the testing of the composite hypothesis that data follow a Pareto distribution with unspecified parameters. Furthermore, note that the sample sizes considered in the two papers are quite distinct; while \cite{chu2019review} considers the performance of tests for larger sample sizes, our focus is on the performance of the tests in the case of smaller samples. Additionally, \cite{chu2019review} employs only maximum likelihood estimation, whereas the current paper uses both maximum likelihood and the adjusted method of moments estimators. In the study presented here, we compare the powers achieved by the various tests using the two different estimation techniques and we demonstrate that the parameter estimation method, perhaps surprisingly, substantially influences the powers associated with the various tests. Lastly, the critical values used in \cite{chu2019review} are obtained using a bootstrap approach; in Section 3, we show that it is possible to obtain critical values independent of the estimated parameters when using maximum likelihood estimation. This allows us to estimate critical values without resorting to a bootstrap procedure in the case where maximum likelihood parameter estimates are employed.

In order to proceed we introduce some notation. Let $X,X_1,X_2,...,X_n$ be independent and identically distributed (i.i.d.) continuous positive random variables with an unknown distribution function $F$. Let $X_{(1)}\leq X_{(2)}\leq... \leq X_{(n)}$ denote the order statistics of $X_1,X_2,...,X_n$. Denote the Pareto distribution function by
\begin{equation}
  F_{\beta,\sigma}(x)=1-
    \left(\frac{x}{\sigma}\right)^{-\beta},\quad x \geq \sigma, \label{ParDist}
\end{equation}
and the density function by
\begin{equation*}
  f_{\beta,\sigma}(x)=\frac{\beta\sigma^\beta}{x^{\beta+1}},\quad x \geq \sigma,
\end{equation*}
where $\beta>0$ is a shape parameter and $\sigma>0$ is a scale parameter. To indicate that the distribution of a random variable $X$ is the Pareto distribution with shape and scale parameters $\beta$ and $\sigma$, we make use of the following shorthand notation: $X \sim P(\beta,\sigma)$.

The hypothesis to be tested is that an observed data set is realised from a Pareto distribution, but we distinguish between two distinct hypothesis testing scenarios. In the first scenario, the value of $\sigma$ in (\ref{ParDist}) is \emph{known} while the value of $\beta$ is unspecified. Note that $\sigma$ determines the support of the Pareto distribution. As a result, if the support of the distribution is known, then the value of $\sigma$ is also known. As a concrete example, consider the case of an insurance company. Typically, an insurance claim is subject to a so-called excess, meaning that the insurance company will only receive a claim if it exceeds a known, fixed value. A closely related example is considered in Section 5; here the monetary expenses (above a certain threshold) resulting from wind related catastrophes are examined. Another example is found in \cite{Arn}, where the lifetime tournament earnings of professional golfers are considered. However, only golfers with a total lifetime earning exceeding \$700 000 are considered. In the second hypothesis testing scenario considered, we may be interested in modelling a phenomenon for which the support of $F$ is unknown and the values of both $\beta$ and $\sigma$ require estimation. In both testing scenarios, we are interested in testing the composite goodness-of-fit hypothesis
\begin{equation}\label{H0main}
  H_0: F(x) = F_{\beta,\sigma}(x),
\end{equation}
for some $\beta>0$, $\sigma>0$ and all $x>\sigma$. This hypothesis is to be tested against general alternatives.

The remainder of the paper is organised as follows. Section 2 provides an overview of a large number of tests for the Pareto distribution based on a wide range of characterisations of this distribution. Section 3 considers two types of estimators for the parameters of the Pareto distribution; the method of maximum likelihood as well as a method closely related to the method of moments. This section also details the estimation of critical values for the tests considered. An extensive Monte Carlo study is presented in Section 4. This section investigates and compares the finite sample performance of the tests in various of settings. Section 5 presents a practical implementation of the goodness-of-fit tests as well as the parameter estimation techniques considered. These techniques are demonstrated using a data set comprised of the monetary expenses resulting from wind related catastrophes in 40 separate instances during 1977. Some conclusions are presented in Section 6.

\section{Goodness-of-fit tests for the Pareto distribution}

We discuss various goodness-of-fit tests for the Pareto distribution below; tests are grouped according to the characteristic of the Pareto distribution that the tests are based on. We consider tests utilising the empirical distribution function, likelihood ratios, entropy, phi-divergence, empirical characteristic function as well as Mellin transform. Additionally, the discussion below includes tests based on the so-called inequality curve as well as various characterisations of the Pareto distribution.

\subsection{Tests based on the empirical distribution function (edf)}

Classical edf-based tests, such as the Kolmogorov-Smirnov, Cram\'{e}r-von Mises, and Anderson-Darling tests are based on a distance measure between parametric and non-parametric estimates of the distribution function. The non-parametric estimate of the distribution function of $X_1,X_2,...,X_n$ used is the edf,
\begin{equation*}
    F_n(x) = \frac{1}{n}\sum_{j=1}^{n}I(X_j\leq{x}),
\end{equation*}
with $I(\cdot)$ the indicator function, while the parametric estimate of the distribution function is
\begin{equation*}
  F_{\widehat{\beta}_n,\widehat{\sigma}_n}(x)=1-
    \left(\frac{x}{\widehat{\sigma}_n}\right)^{-\widehat{\beta}_n},
\end{equation*}
where $\widehat{\beta}_n$ and $\widehat{\sigma}_n$ are estimates of the shape and scale parameters of the Pareto distribution. Parameter estimation is discussed in Section 3.

The Kolmogov-Smirnov test statistic, corresponding to the supremum difference between $F_{\widehat{\beta}_n,\widehat{\sigma}_n}$ and $F_n$, is
 \begin{equation*}
    KS_n = \sup_{x\geq \widehat{\sigma}_n}|F_n(x)-F_{\widehat{\beta}_n,\widehat{\sigma}_n}(x)|,
\end{equation*}
which, after some standard calculations, simplifies to 
\begin{equation*}
    KS_n = \max\left\{KS^{+},KS^{-}\right\},
\end{equation*}   
where
\begin{equation*}
    KS^{+} = \max_{1\leq j\leq n}\left[\frac{j}{n}-F_{\widehat{\beta}_n,\widehat{\sigma}_n}(X_{(j)})\right]
\end{equation*}
and
\begin{equation*}
 KS^{-} = \max_{1\leq j\leq n}\left[F_{\widehat{\beta}_n,\widehat{\sigma}_n}(X_{(j)})-\frac{j-1}{n}\right].
\end{equation*}


The remaining edf test statistics considered are $L2$ distances and have the following general form,
\begin{equation}
    n\int_{-\infty}^{\infty}\left[F_n(x)-F_{\widehat{\beta}_n,\widehat{\sigma}_n}(x)\right]^2w(x)\mathrm{d}F_{\widehat{\beta}_n,\widehat{\sigma}_n}(x), \label{gh}
\end{equation} 
where $w(x)$ is some weight function. Choosing $w(x)=1$ in \eqref{gh}, we have the Cram\'{e}r-von Mises test with direct calculable form
\begin{equation*}
    CM_n = \frac{1}{12n} +\sum_{j=1}^{n}\left[F_{\widehat{\beta}_n,\widehat{\sigma}_n}(X_{(j)})-\frac{2j-1}{2n}\right]^2.
\end{equation*}
When choosing $w(x)=\left[F_{\widehat{\beta}_n,\widehat{\sigma}_n}(x)\{1-F_{\widehat{\beta}_n,\widehat{\sigma}_n}(x)\}\right]^{-1}$, we obtain the Anderson-Darling test
\begin{equation*}
    AD_n = -n-\frac{1}{n}\sum_{j=1}^{n}(2j-1)\left[\log\left(F_{\widehat{\beta}_n,\widehat{\sigma}_n}(X_{(j)})\right)+\log\left(1-F_{\widehat{\beta}_n,\widehat{\sigma}_n}(X_{(n+1-j)})\right)\right].
\end{equation*}
Finally, setting $w(x)=[1-F_{\widehat{\beta}_n,\widehat{\sigma}_n}(x)]^{-2}$, we arrive at the so-called modified Anderson-Darling test
\begin{equation*}
  MA_n = \frac{n}{2}-2\sum_{j=1}^{n}F_{\widehat{\beta}_n,\widehat{\sigma}_n}(X_j)-\sum_{j=1}^{n}\left[2-\frac{2j-1}{n}\right]\log\left(1-F_{\widehat{\beta}_n,\widehat{\sigma}_n}(X_{(j)})\right).
\end{equation*}

While the $CM_n$, $AD_n$ and $MA_n$ tests are all $L2$ distances between the parametric and non-parametric estimates of the distribution function, the weight functions used vary the importance allocated to different types of deviations between these estimates. For example, when comparing the Cram\'{e}r-von Mises and Anderson-Darling tests, differences in the tail of the distribution are more heavily weighted in the case of the latter than the former.
For further discussions on these edf-based tests, see, \cite{klar2001goodness} and \cite{d1986tests}. All of the above tests reject the null hypothesis for large values of the test statistics.

\subsection{Tests based on likelihood ratios}

\cite{zhang2002powerful} proposes two general test statistics which are used to test for normality; below we adapt these tests in order to test for the Pareto distribution. The test statistics are of the form
\begin{equation}
    T_n= \int_{-\infty}^{\infty}G_n(x)\mathrm{d}w(x),
\end{equation}
where $G_n(x)$ is the likelihood ratio statistic defined as 
\begin{equation*}
    G_n(x)=2n\left\{F_n(x)\log\left(\frac{F_n(x)}{F_{\widehat{\beta}_n,\widehat{\sigma}_n}(X_{(j)})}\right)+[1-F_n(x)]\log\left(\frac{1-F_n(x)}{1-F_{\widehat{\beta}_n,\widehat{\sigma}_n}(X_{(j)})}\right)\right\}.
\end{equation*}
The two choices of $\mathrm{d}w(x)$ that  \cite{zhang2002powerful} proposes, as well as the test statistics resulting from each of these choices, are presented below. The results are obtained upon setting $F_n(X_{(j)})=(j-\frac{1}{2})/n$.
\begin{itemize}
    \item Choosing $\mathrm{d}w(x)=\left [F_n(x)\{1-F_n(x)\}\right]^{-1}\mathrm{d}F_n(x)$ leads to
\begin{equation*}
    ZA_n=-\sum_{j=1}^{n}\left\{\frac{\log\left(F_{\widehat{\beta}_n,\widehat{\sigma}_n}(X_{(j)})\right)}{n-j+\frac{1}{2}}+\frac{\log\left(1-F_{\widehat{\beta}_n,\widehat{\sigma}_n}(X_{(j)})\right)}{j-\frac{1}{2}}\right\}.
\end{equation*}
\item Choosing $\mathrm{d}w(x)=\left [F_{\widehat{\beta}_n,\widehat{\sigma}_n}(x)\{1-F_{\widehat{\beta}_n,\widehat{\sigma}_n}(x)\}\right]^{-1}\mathrm{d}F_{\widehat{\beta}_n,\widehat{\sigma}_n}(x)$
results in
\begin{equation*}
    ZB_n=\sum_{j=1}^{n}\left\{\log\left(\frac{\left(F_{\widehat{\beta}_n,\widehat{\sigma}_n}(X_{(j)})\right)^{-1}-1}{(n-\frac{1}{2})/(j-\frac{3}{4})-1}\right)\right\}^2.
\end{equation*}
\end{itemize}
Motivated by the high powers often obtained using the modified Anderson-Darling test, we also include the choice $\mathrm{d}w(x)=\{1-F_n(x)\}^{-2}\mathrm{d}F_n(x)$, which leads to the test statistic
\begin{equation*}
    ZC_n=2\sum_{j=1}^{n}\left\{\frac{n(j-\frac{1}{2})}{(n-j+\frac{1}{2})^2}\log\left(\frac{j-\frac{1}{2}}{nF_{\widehat{\beta}_n,\widehat{\sigma}_n}(X_{(j)})}\right)+\frac{n}{n-j+\frac{1}{2}}\log\left(\frac{n-j+\frac{1}{2}}{n(1-F_{\widehat{\beta}_n,\widehat{\sigma}_n}(X_{(j)}))}\right)\right\}.
\end{equation*}
All three of these tests reject the null hypothesis for large values of the test statistics.

Building on the tests for the assumption of normality that \cite{zhang2002powerful} proposes, \cite{noughabi2015testing} adapts two of these test to test the assumption of exponentiality. Neither \cite{zhang2002powerful} nor \cite{noughabi2015testing} derive the asymptotic properties of these tests and rather present extensive Monte Carlo studies to investigate their finite sample performances. The authors found that these tests are  quite powerful compared to other tests (especially the traditional edf-based tests) against a range of alternatives.
\begin{remark*}
\cite{zhang2002powerful} also considers the test
\begin{align*}
    ZD_n=&\sup_{x\in \mathbb{R}}G(x) = \max_{1\leq j\leq n} G(X_{(j)})\\
    =& \max_{1\leq j\leq n} \left\{\left(j-\frac{1}{2}\right) \log\left(\frac{j-\frac{1}{2}}{nF_{\widehat{\beta}_n,\widehat{\sigma}_n}(X_{(j)})} \right)+\left(n-j+\frac{1}{2}\right)\log  \left(\frac{n-j+\frac{1}{2}}{n(1-F_{\widehat{\beta}_n,\widehat{\sigma}_n}(X_{(j)}))}\right)  \right\}.
\end{align*}
However, we do not include $ZD_n$ in our Monte Carlo study as $ZA_n$ and $ZB_n$ proved more powerful in the papers mentioned.
\end{remark*}

\subsection{Tests based on entropy}

A further class of tests is based on the concept of entropy, first introduced in \cite{shannon1948mathematical}.
The entropy of a random variable $X$ with density and distribution functions $f$ and $F$, respectively, is defined to be
\begin{equation}
    H= -\int_{0}^{\infty}f(x)\log (f(x))\mathrm{d}x = \int_{0}^{1}\log\left(\frac{\mathrm{d}}{\mathrm{d}p}F^{-1}(p)\right)\mathrm{d}p,\label{eqh}
\end{equation}
where $F^{-1}(\cdot)$ denotes the quantile function of $X$.
The concept of entropy has been applied in several studies, see, e.g., \cite{kullback1997information},  \cite{kapur1994measures} and \cite{vasicek1976test}, where, in particular, \cite{vasicek1976test} proposes using
\begin{equation}
    H_{n,m}=\frac{1}{n}\sum_{j=1}^{n}\log\left\{\left(\frac{n}{2m}\right)(X_{(j+m)}-X_{(j-m)})\right\} \label{eqhmn}
\end{equation}
as an estimator for $H$, where $X_{(j)}=X_{(1)}$ for $j<1$, $X_{(j)}=X_{(n)}$ for $j>n$, and $m$ is a window width subject to $m\leq \frac{n}{2}$. We now consider two goodness-of-fit tests based on concepts related to entropy: the Kullback-Leibler divergence and the Hellinger distance, where $H$ is estimated by $H_{n,m}$ in the test statistic.

The Kullback-Leibler divergence between any arbitrary density function, $f$, and the Pareto density, $f_{\beta,\sigma}$, is defined to be \citep[see, e.g.,][]{kullback1997information} 
\begin{equation*}
     KL=-\int_{0}^{\infty} f(x)\log\left(\frac{f(x)}{f_{\beta,\sigma}(x)}\right)\mathrm{d}x. 
\end{equation*}
It follows that the Kullback-Leibler divergence can also be express in terms of entropy:
\begin{equation}
   KL= -H -\int_{0}^{\infty} f(x)\log(f_{\beta,\sigma}(x))\mathrm{d}x.\label{hh}
\end{equation} 
Estimating \eqref{hh} by the empirical quantities mentioned above, we obtain the test statistic
\begin{equation*}
   KL_{n,m}= -H_{n,m}-\log\left(\widehat{\beta}_n\right)-\widehat{\beta}_n\log\left(\widehat{\sigma}_n\right)+\left(\widehat{\beta}_n+1\right)\frac{1}{n}\sum_{j=1}^{n}\log(X_j).
\end{equation*}
This test rejects the null hypothesis for large values of $KL_{n,m}$. \cite{ahrari2019goodness} uses a similar test statistic in order to test the goodness-of-fit hypothesis for the Rayleigh distribution and proves that the test is consistent. In their simulation study the authors find that the test compared favourably to other competing tests.

The Hellinger distance between two densities $f$ and $f_{\beta,\sigma}$ is defined as \citep[see, e.g.,][]{jahanshahi2016goodness} 
 \begin{equation*}
   HD = \frac{1}{2}\int_{0}^{\infty}\left(\sqrt{f(x)}-\sqrt{f_{\beta,\sigma}(x)}\right)^2\mathrm{d}x.
\end{equation*}
By setting $F(x)=p$, the Hellinger distance can be expressed in terms of the quantile function as follows
 \begin{equation*}
   HD = \frac{1}{2}\int_{0}^{1}\left(\sqrt{\left(\frac{\mathrm{d}}{\mathrm{d}p}F^{-1}(p)\right)^{-1}}-\sqrt{\frac{\beta\sigma^\beta}{(F^{-1}(p))^{\beta+1}}}\right)^2\frac{\mathrm{d}}{\mathrm{d}p}F^{-1}(p)\mathrm{d}p.
\end{equation*}
From \eqref{eqh} and \eqref{eqhmn} it can be argued that $\frac{\mathrm{d}}{\mathrm{d}p}F^{-1}(p)$ can be estimated by $\{\frac{n}{2m}(X_{(j+m)}-X_{(j-m)})\}$. The resulting test statistic is given by
\begin{equation*}
   HD_{n,m} =\frac{1}{2n}\sum_{j=1}^{n}
   \frac
   {\left[{\left\{\frac{n}{2m}(X_{(j+m)}-X_{(j-m)})\right\}^{-1/2}}-\left(f_{\widehat{\beta}_n,\widehat{\sigma}_n}(X_j)\right)^\frac{1}{2}\right]^2}
   {\left\{\frac{n}{2m}(X_{(j+m)}-X_{(j-m)})\right\}^{-1}}.
\end{equation*}
This test rejects the null hypothesis for large values of $HD_{n,m}$.

\cite{jahanshahi2016goodness} uses similar arguments to propose a goodness-of-fit test for the Rayleigh distribution and proves that the test is consistent in that setting. In addition, they also propose a method for obtaining the optimum value of $m$ by minimising bias and mean square error (MSE). In a finite sample power comparison, \cite{jahanshahi2016goodness} finds that $HD_{n,m}$ produces the highest estimated powers against the majority of alternatives considered. In the case of alternatives considered with non-monotone hazard rates, the entropy-based tests outperform the remaining tests by some margin.

\subsection{Tests based on the phi-divergence}

The phi-divergence between an arbitrary density, $f$, and $f_{\beta, \sigma}$ is
\begin{equation*}
    D_\phi(f,f_{\beta, \sigma})=\int_{\sigma}^{\infty}\phi\left(\frac{f(x)}{f_{\beta, \sigma}(x)}\right)f_{\beta, \sigma}(x)\mathrm{d}x,
\end{equation*}
where $\phi: [0,\infty)\longrightarrow (-\infty,\infty)$ is a convex function such that $\phi(1)=0$ and $\phi''(1)>0$. It is further known 
\citep[see, e.g.,][]{choi2006testing,csiszar1967topological}
that if $\phi$ is strictly convex in a neighbourhood of $x=1$, then $D_\phi(f,f_{\beta, \sigma})=0$ if, and only if, $f=f_{\beta, \sigma}$.
\cite{alizadeh2016tests} use this property to construct goodness-of-fit tests for a variety of different distributions.
Let $E_F[\cdot]$ denote an expectation taken with respect to the distribution $F$. By noting that
\begin{equation*}
        D_\phi(f,f_{\beta, \sigma})=\int_{\sigma}^{\infty}\phi\left(\frac{f(x)}{f_{\beta, \sigma}(x)}\right) \frac{f_{\beta, \sigma}(x)}{f(x)}\mathrm{d}F(x) =E_F\left[\phi \left(\frac{f(X)}{f_{\beta, \sigma}(X)}\right) \frac{f_{\beta, \sigma}(X)}{f(X)}\right],
\end{equation*}
it follows that
$D_\phi(f,f_{\beta, \sigma})$ can be estimated by
\begin{equation}
   \widehat{D}_\phi(\widehat{f}_h,f_{\widehat{\beta}_n,\widehat{\sigma}_n})=\frac{1}{n}\sum_{j=1}^{n}\left[\phi \left(\frac{\widehat{f}_h(X_j)}{f_{\widehat{\beta}_n,\widehat{\sigma}_n}(X_j)}\right) \frac{f_{\widehat{\beta}_n,\widehat{\sigma}_n}(X_j)}{\widehat{f}_h(X_j)}\right], \label{star}
\end{equation}
where $\widehat{f}_h(x)=\frac{1}{nh}\sum_{j=1}^{n}k\left(\frac{x-X_j}{h}\right)$ is the kernel density estimator with kernel function $k(\cdot)$ and bandwidth $h$.

In the Monte Carlo study in Section 4, we use the standard normal density function as kernel and choose $h=1.06sn^{-\frac{1}{5}}$, where $s$ is the unbiased sample standard deviation \citep[see, e.g.,][]{silverman2018density}. We will use the following four choices of $\phi$:
\begin{itemize}
    \item The Kullback-Liebler distance (DK) with $\phi(x)=x\log (x)$.
    \item The Hellinger distance (DH) with $\phi(x)=\frac{1}{2}(\sqrt{x}-1)^2$.
    \item The Jeffreys divergence distance (DJ) with $\phi(x)=(x-1)\log (x)$.
    \item The total variation distance (DT) with $\phi(x)=|x-1|$.
\end{itemize}
A variety of test statistics can be constructed from (\ref{star}) using the above choices of $\phi$. The test statistics corresponding to these choices are
\begin{eqnarray*}
    DK_n&=& \frac{1}{n}\sum_{j=1}^{n}\left[\log\left(\frac{\widehat{f}_h(X_j)}{f_{\widehat{\beta}_n,\widehat{\sigma}_n}(X_j)}\right)\right],\\
    DH_n&=& \frac{1}{2n}\sum_{j=1}^{n}\left[\left(1-\sqrt{\frac{\widehat{f}_h(X_j)}{f_{\widehat{\beta}_n,\widehat{\sigma}_n}(X_j)}}\right)^2 \frac{f_{\widehat{\beta}_n,\widehat{\sigma}_n}(X_j)}{\widehat{f}_h(X_j)}\right],\\
    DJ_n&=& \frac{1}{n}\sum_{j=1}^{n}\left[\left(1-   \frac{f_{\widehat{\beta}_n,\widehat{\sigma}_n}(X_j)}{\widehat{f}_h(X_j)}\right)\log\left(\frac{\widehat{f}_h(X_j)}{f_{\widehat{\beta}_n,\widehat{\sigma}_n}(X_j)}\right)\right], \text{ and} \\
    DT_n&=& \frac{1}{n}\sum_{j=1}^{n}\left[\left|\frac{\widehat{f}_h(X_j)}{f_{\widehat{\beta}_n,\widehat{\sigma}_n}(X_j)}-1\right|\frac{f_{\widehat{\beta}_n,\widehat{\sigma}_n}(X_j)}{\widehat{f}_h(X_j)}\right].
\end{eqnarray*}
All tests reject the null hypothesis for large values of the test statistics.

In addition to showing that the tests above are consistent against fixed alternatives, \cite{alizadeh2016tests} also uses $DK_n$, $DH_n$, $DJ_n$ and $DT_n$ to test the goodness-of-fit hypothesis for the normal, exponential, uniform and Laplace distributions. The Monte Carlo study included in \cite{alizadeh2016tests} indicates that $DK_n$ produces the highest powers amongst the phi-divergence type tests. When comparing the performance of these tests, the powers associated with $DK_n$ were higher than the others. As a result, only $DK_n$ is included in the Monte Carlo study presented in Section 4.

\subsection{A test based on the empirical characteristic function}

A large number of goodness-of-fit tests have been developed for a variety of distributions based on empirical characteristic functions
\citep[see, e.g.,][]{klar2005tests,meintanis2009goodness,betsch2020testing}.
For a review of testing procedures based on the empirical characteristic functions see, e.g., \cite{meintanis2016review}.

Recall that the characteristic function (cf) of a random variable $X$ with distribution $F_{\theta}$ is given by
\begin{equation*}
    \varphi_{\theta}(t)=E[\mathrm{e}^{itX}]=\int \mathrm{e}^{itx} \mathrm{d}F_{\theta}(x),
\end{equation*}
with $i=\sqrt{-1}$ the imaginary unit. The empirical characteristic function (ecf) is defined to be
\begin{equation*}
    \varphi_n(t)=\frac{1}{n}\sum_{j=1}^{n}\mathrm{e}^{itX_j}.
\end{equation*}

As a general test statistic, one can use a weighted $L2$ distance between the fitted cf under the null hypothesis and the ecf,
\begin{equation*}
    \int_{-\infty}^{\infty}|\varphi_n(x)-\varphi_{\widehat{\theta}}(x)|^2w(x)\mathrm{d}x,
\end{equation*}
where $\widehat{\theta}$ represents the estimated values of the parameters of the hypothesised distribution and $w(\cdot)$ is a suitably chosen weight function ensuring that the integral is finite. Commonly used choices for the weight function are $w(x)=\mathrm{e}^{-a|x|}$ and $w(x)=\mathrm{e}^{-ax^2}$, respectively derived from the kernels of the Laplace and normal density functions, where $a>0$ is a user defined tuning parameter.

The characteristic function of the Pareto distribution does not admit a closed form expression, making $T_n$ intractable irrespective of the choice of the weight function. In order to circumvent this problem, we use the test proposed in \cite{meintanis2009goodness}. In order to perform this test, the data are transformed so as to approximately follow a standard uniform distribution under the null hypothesis. The test statistic used is a weighted $L2$ distance between the ecf of the transformed data and the cf of the standard uniform distribution.

In order to specify the test statistic, let
\begin{equation*}
    \widehat{U}_j=F_{\widehat{\beta}_n,\widehat{\sigma}_n}(X_j), j=1,...,n,
\end{equation*}
be the transformed data and denote by $\widehat{\varphi}_n(x)$ the ecf of the transformed data. Let
\begin{equation*}
    \varphi_U(x)=\frac{\sin (x)+i(1-\cos (x))}{x}
\end{equation*}
be the cf of the uniform distribution. \cite{meintanis2009goodness} proposes the test
\begin{equation*}
    S_{n,a}=\int_{-\infty}^{\infty}|\varphi_U(x)-\widehat{\varphi}_n(x)|^2w(x)\mathrm{d}x.
\end{equation*}
Upon setting $w(x)=\mathrm{e}^{-a|x|}$, $S_{n,a}$ simplifies to
\begin{eqnarray*}
    S_{n,a}&=&\frac{1}{n}\sum_{j,k=1}^{n}\frac{2a}{(\widehat{U}_j-\widehat{U}_k)^2+a^2}+2n\left[2\tan^{-1}\left(\frac{1}{a}\right)-a\log\left(1+\frac{1}{a^2}\right)\right]\\
    && -\, 4\sum_{j=1}^{n}\left[\tan^{-1}\left(\frac{\widehat{U}_j}{a}\right)+\tan^{-1}\left(\frac{1-\widehat{U}_j}{a}\right)\right].
\end{eqnarray*}
The test rejects the null hypothesis for large values of the test statistic.
Although \cite{meintanis2009goodness} does not explicitly use the resulting statistic to test for the Pareto distribution, it is demonstrated that this test is competitive when testing for the gamma, inverse Gaussian, and normal distributions.

\subsection{A test based on the Mellin transform}

\cite{meintanis2009unified} introduces a test based on the moments of the reciprocal of the random variable $X$. If $X$ follows a Pareto distribution, then $E(X^t)$, $t>0$, only exists when $t<\beta$. On the other hand, the Mellin transform of $X$, given by
\begin{equation*}
    M(t)=E(X^{-t}), t>0,
\end{equation*}
exists for all $t>0$ if $X$ is a Pareto random variable. Given an observed sample, the empirical Mellin transform is defined to be
\begin{equation*}
    \widehat{M}_n(t)=\frac{1}{n}\sum_{j=1}^{n} X_j^{-t}.
\end{equation*}

If $X$ is a $P(\beta,\sigma)$ random variable, then $M(t)$ satisfies \citep[see, e.g.,][]{meintanis2009unified}
\begin{equation*}
    D(t)=(\beta+t)\sigma^{t}M(t)-\beta=0, \quad t>0.
\end{equation*}
Based on a random sample, $D(t)$ can be estimated by
\begin{equation*}
    D_n(t)=(\widehat{\beta}_n+t)\widehat{M}_n(t)-\widehat{\beta}_n.
\end{equation*}
\cite{meintanis2009unified} proposes a weighted $L2$ distance between $D_n(t)$ and $0$ as test statistic;
\begin{equation*}
    G_{n,w}=n\int_{0}^{\infty}D_n^2(t)w(t)\mathrm{d}t,
\end{equation*}
where $w(t)$ is a suitable weight function. After some algebra $G_{n,w}$ simplifies to
\begin{eqnarray*}
    G_{n,w}&=&\frac{1}{n}\left[(\widehat{\beta}_n+1)^2\sum_{j,k=1}^{n}I^{(0)}_w({X_jX_k})+\sum_{j,k=1}^{n}I^{(2)}_w({X_jX_k})+2(\widehat{\beta}_n+1)\sum_{j,k=1}^{n}I^{(1)}_w({X_jX_k})\right]\\
    &&+\widehat{\beta}_n\left[n\widehat{\beta}_nI^{(0)}_w(1)-2(\widehat{\beta}_n+1)\sum_{j=1}^{n}I^{(0)}_w({X_j})-2\sum_{j=1}^{n}I^{(1)}_w({X_j})\right],
   \end{eqnarray*}
where 
\begin{equation*}
    I^{(m)}_w(t)=\int_{0}^{\infty}(t-1)^m\frac{1}{x^t}w(t)\mathrm{d}t, \quad m=0,1,2.
\end{equation*}
Choosing $w(x)=\mathrm{e}^{-ax}$, one has
\begin{eqnarray*}
    I^{(0)}_a(x)&=&(a+\log x)^{-1},\\
    I^{(1)}_a(x)&=&\frac{1-a-\log x}{(a+\log x)^2},
\end{eqnarray*}
and
\begin{equation*}
    I^{(2)}_a(x)=\frac{2-2a+a^2+2(a-1)\log x +\log^2 x}{(a+\log x)^3},
\end{equation*}
culminating in an easily calculable test statistic, henceforth denoted by $G_n$.

The test rejects the null hypothesis for large values of the test statistic. \cite{meintanis2009unified} proves the consistency of the test against fixed alternatives and uses a Monte Carlo study to demonstrate that the power performance of the test compares favourably to that of the classical goodness-of-fit tests.

\subsection{A test based on an inequality curve}

Let $X$ be a positive random variable with distribution function $F$ and finite mean $\mu$. Let $L(p)=Q(F^{-1}(p))$, with
\[
    F^{-1}(p)=\inf\{x:F(x)\geq p\},
\]
the generalised inverse of $F$ and
\[
    Q(x)=\frac{1}{\mu}\int_{0}^{x}t\mathrm{d}F(t),
\]
the first incomplete moment of $X$. Using this notation, the inequality curve $\lambda(p)$, $p\in(0,1)$ is defined to be \citep[see, e.g.,][]{zenga1984proposta}
\begin{equation*}
    \lambda(p)=1-\frac{\log (1-L(p))}{\log (1-p)}.
\end{equation*}
\cite{taufer2021graphical} proposes a test based on the constant inequality curve exhibited by the $P(\beta,\sigma)$ distribution for some $\sigma>0$. \cite{taufer2021graphical} proves the following characterisation for the Pareto distribution based on $\lambda (p)$.
\begin{theorem}
    The inequality curve $\lambda (p)$ is equal to $\frac{1}{\beta}$ over all values of $p$, $p\in(0,1)$ if, and only if, $F$ is the Pareto distribution function, $F_{\beta,\sigma}$.
\end{theorem}

In order to use this characterisation to develop goodness-of-fit tests, \cite{taufer2021graphical} uses the following approach. Defining the empirical version of $Q(x)$ as
\begin{equation*}
    Q_n(x)=\frac{\sum_{j=1}^{n}X_jI(X_j\leq{x})}{\sum_{j=1}^{n}X_j},
\end{equation*}
the estimator for $L(p)$ becomes
\begin{equation*}
    L_n(p)=Q_n(F_n^{-1}(p))=\frac{\sum_{j=1}^{i}X_{(j)}}{\sum_{k=1}^{n}X_{(k)}}, \quad \frac{j-1}{n}\leq p \leq \frac{j}{n}, \quad j=1,2,\dots,n,
\end{equation*}
where $F_n^{-1}(p)=\inf\{x:F_n(x)\leq p\}$. Finally, an estimator for $\lambda (p)$ is given by
\begin{equation*}
    \widehat{\lambda}_j=1-\frac{\log (1- L_n(p_j))}{\log (1-p_j)}, \quad j=1,2,...,n-\lfloor\sqrt{n}\rfloor, \quad p_j=\frac{j}{n}.
\end{equation*}

The choice $j=1,2,\dots,n-\lfloor\sqrt{n}\rfloor$ ensures that $\widehat{\lambda}_j$ is a consistent estimator for $\lambda$  \citep[see][]{taufer2021graphical}. 
Setting $m=n-\lfloor\sqrt{n}\rfloor$, Theorem 2.1 states that under the null hypothesis, for any choice of $p_j$, $0<p_j<1$, $j=1,2,\dots,m$, and $\beta>1$, we have the linear equation
\begin{equation*}
    \lambda_j=\beta_0+\beta_1p_j,
\end{equation*}
with $\beta_0=\frac{1}{\beta}$ and $\beta_1=0$.
Now, based on the data $X_1,...,X_n$, we can obtain estimators for $\beta_0$ and $\beta_1$ from the regression
\begin{equation*}
   \widehat{\lambda}_j=\beta_0+\beta_1p_j+\varepsilon_j,
\end{equation*}
where $\varepsilon_j=\widehat{\lambda}_j-\lambda_j$.

The least squares estimators for $\beta_0$ and $\beta_1$ are given by
\begin{equation*}
   \widehat{\beta}_0=\frac{1}{m}\sum_{j=1}^{m} \widehat{\lambda}_j
\text{\quad and \quad} 
   \widehat{\beta}_1=\sum_{j=1}^{m}\frac{\widehat{\lambda}_j(p_j-\Bar{p})}{S_p^2},
\end{equation*}
where $\Bar{p}=\frac{m(m+1)}{2n^2}$ and $S_p^2=\frac{m(m^2-1)}{12n^2}$.

Testing the hypothesis in is equivalent to testing the hypothesis
\begin{equation*}
   H_0:\beta_1=0
\text{\quad versus \quad}
   H_A:\beta_1\ne0,
\end{equation*}
where the null hypothesis is rejected for large values of $|\widehat{\beta}_1|$. In a finite sample study, \cite{taufer2021graphical} finds that this test is oversized in the case of small sample sizes ($n=20$), but achieves the nominal significance level for larger samples ($n=100$). The results indicate that the test compares favourably against the traditional tests in terms of estimated powers. Since the focus of the current research is on the finite sample performance of tests in the case of small samples, we do not include this test in the Monte Carlo study presented in Section 4.

\subsection{Tests based on various characterisations of the Pareto distribution}

A wide range of characterisations of the Pareto distribution is available and several have been used to develop goodness-of-fit tests. In what follows, we state these characterisations and discuss the associated test in each case. It should be noted that, although the tests below are equally useful in the situation where both parameters of the Pareto distribution are required to be estimated, the asymptotic theory was developed in the setting where the scale parameter is known.

Each of the subsections below are dedicated to one of these characterisations and contains a brief discussion on the associated test.

\subsubsection{Characterisation 1 \texorpdfstring{\citep{obradovic2015goodness}}{}}

\textit{Let $X$ and $Y$ be i.i.d. positive absolutely continuous random variables. The random variable $X$ and $\max \left\{\frac{X}{Y}, \frac{Y}{X}\right\}$ have the same distribution if, and only if, $X$ follows a Pareto distribution}.

\cite{obradovic2015goodness} provides the proof for this characterisation and proposes two test statistics based on it. In order to specify these test statistics, denote by
\begin{equation*}
   M_n(x) = \binom{n}{2}^{-1} \ \sum_{i=1}^{n-1}\sum_{j=i+1}^{n}I\left\{\max\left(\frac{X_i}{X_j},\frac{X_j}{X_i}\right) \leq x\right\}, \quad x\geq1,
\end{equation*}
the $U$-empirical distribution function of the random variable $\mathrm{max}\{X/Y,Y/X\}$. The test statistics are specified to be
\begin{equation*}
    T_n = \int_{1}^{\infty}\left[M_n(x)-F_n(x)\right]\mathrm{d}F_n(x)
\end{equation*}
and
\begin{equation*}
    V_n = \sup_{x\geq1}|M_n(x)-F_n(x)|.
\end{equation*}
Both of these tests reject the null hypothesis for large values of the test statistics. \cite{obradovic2015goodness} calculates Bahadur efficiencies for selected alternative distributions and also determines some of the locally optimal alternatives. The mentioned paper also derives the null distribution of $T_n$ and shows that $\sqrt{n}T_n$ converges to a centered normal random variable with variance $5/108$. 
A limited Monte Carlo study shows that the tests $T_n$ and $V_n$ are competitive against the traditional $KS_n$ and $CM_n$ tests.

\subsubsection{Characterisation 2 \texorpdfstring{\citep{allison2022distribution}}{}}

\textit{Let $X, X_1,...,X_n$ be i.i.d. positive absolutely continuous random variables from some distribution function $F$. The random variables $\sqrt[m]{X}$ and $\min(X_1,...,X_m)$ have the same distribution if, and only if, $F$ is the Pareto distribution, for all integers $2\leq m\leq n$}.


Using $m$ as a tuning parameter, \cite{allison2022distribution} proposes three classes of tests for the Pareto distribution based on the characterisation above. The test statistics used are discrepancy measures between the empirical distribution of $\min\{X_1,...,X_m\}$ and the $V$-empirical distribution of $\sqrt[m]{X}$, defined as
\begin{equation*}
    \Delta_{n,m}(x)=\frac{1}{n}\sum_{j=1}^{n}I\{X_{j}^\frac{1}{m}\leq x\}-\frac{1}{n^m}\sum_{j_1,...,j_m=1}^nI\{\min(X_{j_{1}},...,X_{j_{m}})\leq x\}.
\end{equation*}
Based on $\Delta_{n,m}$, the authors propose the following test statistics;
\begin{eqnarray*}
    I_{n,m} &=& \int_{1}^{\infty}\Delta_{n,m}(x) \mathrm{d}F_n(x), \\
    K_{n,m} &=& \sup_{x\geq1}|\Delta_{n,m}(x)|, \\
    M_{n,m} &=& \int_{1}^{\infty}\Delta_{n,m}^2(x) \mathrm{d}F_n(x). \\
\end{eqnarray*}
All three tests reject the null hypothesis for large values of the test statistics. \cite{allison2022distribution} investigates the asymptotic properties of these tests where, upon calculating and comparing the Bahadur efficiencies, it is determined that the test $I_{n,m}$ has the best performance among the three in terms of local efficiency. This result is reinforced by a finite sample power study which results in the recommendation of choosing $I_{n,m}$ with $m=2$.

\subsubsection{Characterisation 3 \texorpdfstring{\citep{rossberg1972characterization}}{}}

\cite{obradovic2015asymptotic} uses the following  special case of Rossberg's characterisation of the Pareto distribution to construct a goodness-of-fit test: 

\textit{Let $X_1$, $X_2$, and $X_3$ be i.i.d. positive absolutely continuous random variables and denote the corresponding order statistics by $X_{(1)} \leq X_{(2)} \leq X_{(3)}$. If $X_{(2)}/X_{(1)}$ and $\min(X_1,X_2)$ are identically distributed, then $X_1$ follows a Pareto distribution}.

In order to base a test on this characterisation, \cite{obradovic2015asymptotic} suggests estimating the distribution of $X_{(2)}/X_{(1)}$ by
\begin{equation*}
    G_n(x)=\frac{1}{n^3} \sum_{i=1}^{n}\sum_{j=1}^{n}\sum_{k=1}^{n}I \{\text{median}(X_i,X_j,X_k)/\min(X_i,X_j,X_k)\leq x\},\quad x\geq1,
\end{equation*}
and the distribution of $\min(X_1,X_2)$ by
\begin{equation*}
    H_n(x)= \frac{1}{n^2} \sum_{i=1}^{n}\sum_{j=1}^{n}I\{\min(X_i,X_j)\leq x\},\quad x\geq1.
\end{equation*}
Tests can be based on the discrepancy between $G_n$ and $H_n$; \cite{obradovic2015asymptotic} proposes the test statistics
\begin{equation*}
    I_n^{[1]}= \int_{1}^{\infty}\left(G_n(x)-H_n(x)\right) \mathrm{d}F_n(x),
\end{equation*}
and
\begin{equation*}
    D_n^{[1]}= \sup_{x\geq1}|G_n(x)-H_n(x)|.
\end{equation*}
Both tests reject the null hypothesis for large values of the test statistics. Bahadur efficiencies for these tests are presented in \cite{obradovic2015asymptotic} where the results show that, while no test outperforms all others, each test is found to be locally optimal against certain classes of alternatives.

\subsubsection{Characterisation 4 \texorpdfstring{\citep{obradovic2015asymptotic}}{}}

In addition to the tests above, \cite{obradovic2015asymptotic} also proposes tests for the Pareto distribution based on the following characterisation which is linked to a characterisation of the exponential distribution due to \cite{ahsanullah1978characterization}.

\textit{Let $X_1,X_2$ and $X_3$ be i.i.d. positive absolutely continuous random variables with strictly monotone distribution function and monotonically increasing or decreasing hazard function and denote the order statistics by $X_{(1)}\leq X_{(2)}\leq X_{(3)}$. The random variable $X_{(3)}/X_{(2)}$ and $X_{(2)}/X_{(1)}$ have the same distribution if, and only if, the distribution of X follows a Pareto distribution}.

The test statistics that \cite{obradovic2015asymptotic} proposes based on this characterisation are
\begin{equation*}
    I_n^{[2]}= \int_{1}^{\infty}\left(J_n(x)-K_n(x)\right) \mathrm{d}F_n(x), \\
\end{equation*}
and 
\begin{equation*}
    D_n^{[2]}= \sup_{x\geq1}|J_n(x)-K_n(x)|,
\end{equation*}
where
\begin{equation*}
    J_n(x)=\frac{1}{n^3} \sum_{i=1}^{n}\sum_{j=1}^{n}\sum_{k=1}^{n}I\{\max(X_i,X_j,X_k)/\text{median}(X_i,X_j,X_k)\leq x\}, \quad x\geq1, \\
\end{equation*}
and
\begin{equation*}
    K_n(x)= \frac{1}{n^3} \sum_{i=1}^{n}\sum_{j=1}^{n}\sum_{k=1}^{n}I\{\text{median}(X_i,X_j,X_k)/\{\min(X_i,X_j,X_k)\}^2\leq x\}, \quad x\geq1.
\end{equation*}
Both tests reject the null hypothesis for large values of the test statistic. As with the tests given in Section 2.8.3, \cite{obradovic2015asymptotic} concludes that, while neither of the tests was dominant against all alternatives in terms of local efficiency, they are both locally optimal for certain classes of alternatives.


\section{Parameter and critical value estimation}

In this section we discuss two popular methods for the estimation of the parameters of the Pareto distribution: the method of maximum likelihood  as well as a method closely related to moment matching. The empirical results in Section 4 demonstrate that the choice of estimation method used has a profound effect on the powers achieved by the tests considered. As a result, it is necessary to discuss the procedures in some detail.

We consider parameter estimation in the setting where both $\beta$ and $\sigma$ are required to be estimated. In the testing scenario where $\sigma$ is known, the estimated value of $\sigma$ can simply be replaced by this known value. 

For each estimation method we also discuss how the critical values are estimated.

\subsection{Maximum likelihood estimators (MLEs)}

In the case where both $\sigma$ and $\beta$ are unknown, the MLEs of $\sigma$ and $\beta$ are respectively given by
\begin{equation*}
    \widehat{\sigma}_n: = \widehat{\sigma}(X_1,...,X_n) = X_{(1)},
\end{equation*}
and
\begin{equation*}
    \widehat{\beta}_n: = \widehat{\beta}(X_1,...,X_n) = \frac{n}{\sum_{j=1}^{n}\log\left(\frac{X_j}{\widehat{\sigma}_n}\right)}.
\end{equation*}
Note that if we transform $X_1,...,X_n$ as follows:
\begin{equation}
    Y_j = \left(\frac{X_j}{ \widehat{\sigma}_n}\right)^{\widehat{\beta}_n}, \ j = 1,...,n,\label{ag}
\end{equation} 
then
\begin{equation*}
    \widehat{\sigma}(Y_1,...,Y_n) =1,
\end{equation*}
and
\begin{equation*}
    \widehat{\beta}(Y_1,...,Y_n) = \frac{n}{\sum_{j=1}^{n}\log\left(\frac{Y_j}{Y_{(1)}}\right)} = \frac{n}{\widehat{\beta}(X_1,...,X_n)\sum_{j=1}^{n}\log\left(\frac{X_j}{X_{(1)}}\right)} = \frac{ \widehat{\beta}(X_1,...,X_n)}{ \widehat{\beta}(X_1,...,X_n)} = 1.
\end{equation*}

As can be seen above, the transformation in $\eqref{ag}$ ensures that, when the Pareto distribution is fitted to $Y_1,...,Y_n$, the resulting parameter estimates are fixed at $\widehat{\sigma}_n = \widehat{\beta}_n = 1$. This enables us to approximate fixed critical values by Monte Carlo simulations not depending on $\widehat{\sigma}_n$ or $\widehat{\beta}_n$. As a result, the limit null distribution is independent of the values of $\sigma$ and $\beta$ if the data are transformed as in $\eqref{ag}$. This result is quite surprising as it essentially renders the critical values for tests for the Pareto distribution shape invariant in the case where estimation is performed using MLE. We are not aware of any other distribution containing a shape parameter for which this can be accomplished.

It should be noted that, if the transformation in $\eqref{ag}$ is used, then the sample minimum is $Y_{(1)}=1$. This leads to computational issues for several of the tests discussed above. Specifically, the calculation of $AD_n$, $ZA_n$, $ZB_n$ and $ZC_n$ break down. In order to circumvent these numerical problems, we set $Y_{(1)}=1.0001$ when computing these test statistics.

\begin{remark*}
    The test proposed in \cite{taufer2021graphical}, see Section 2.7, assumes that the mean of the Pareto distribution fitted to the transformed values is finite. Let $\widehat\mu_n:=\widehat\beta_n/(\widehat\beta_n-1)$ denote the mean of the fitted Pareto distribution; $\widehat\mu_n$ is finite if, and only if, $\widehat\beta_n>1$. As a result, the transformation in \eqref{ag} leads to numerical problems with the implementation of this test. In order to obtain critical values for this test, we recommend using the transformation $Y_j = \left(\frac{X_j}{ \widehat{\sigma}_n}\right)^{\widehat{\beta}_n/2}, \ j = 1,...,n$, which results in $\widehat\mu:=2$.
\end{remark*}

\subsection{Adjusted method of moments estimators (MMEs)}

The traditional implementation of the method of moments requires that both the mean and the variance of the distribution be finite. In the case of the Pareto distribution, this implies that $\beta>2$. As a result, the traditional method of moments estimators are not consistent when estimating the parameters of a $P(\beta,\sigma)$ distribution when $\beta < 2$.

A partial solution to the problem explained above is found when using the so-called adjusted method of moments estimators proposed in \cite{quandt1964statistical}. Instead of choosing parameter estimates so as to equate the first two population moments to the first two sample moments, \cite{quandt1964statistical} equates the first population and sample moments as well as equating the observed minimum to the expected value of the sample minimum. The resulting estimators are 
\begin{equation*}
    \widetilde{\beta}_n:=\widetilde{\beta}(X_1,...,X_n) =   \frac{n\Bar{X}-X_{(1)}}{n(\Bar{X}-X_{(1)})},
\end{equation*}
and
\begin{equation*}
    \widetilde{\sigma}_n:= \widetilde{\sigma}(X_1,...,X_n) = \frac{\Bar{X}\widetilde{\beta}_n-\Bar{X}}{\widetilde{\beta}_n}.
\end{equation*}
Note that this method only requires the assumption that the population mean is finite, meaning that we assume only that $\beta>1$.

Unlike the case of maximum likelihood, we are unable to obtain fixed critical values; the critical values are functions of the estimated shape parameter $\widetilde{\beta}_n$. We provide the following bootstrap algorithm for the estimation of critical values.
\begin{enumerate}
    \item Based on data $X_1,...,X_n$, estimate $\beta$ and $\sigma$ by $\widetilde{\beta}_n$ and $\widetilde{\sigma}_n$, respectively.
    \item Obtain a parametric bootstrap sample $X_1^*,...,X_n^*$ by sampling independently from $F_{\widetilde{\beta}_n,\widetilde{\sigma}_n}$.
    \item Calculate $\widetilde{\beta}_n^*=\widetilde{\beta}(X_1^*,...,X_n^*)$, $\widetilde{\sigma}_n^*=\widetilde{\sigma}(X_1^*,...,X_n^*)$, and the value of the test statistic say $S^*=S(X_1^*,...,X_n^*)$.
    \item Repeat steps 2 and 3 $B$ times to obtain $S_1^*,...,S_B^*$ and  obtain the order statistics $S_{(1)}^*\leq...\leq S_{(B)}^*$.
    \item The estimated critical value at a $\alpha \times 100\%$ significance level is $\widehat{C}_n=S^*_{(B\lfloor 1-\alpha \rfloor)}$, where $\lfloor x \rfloor$ denotes the floor of $x$.
\end{enumerate}

We now turn our attention to the numerical powers of the tests obtained using the two estimation methods discussed above.

\section{Monte Carlo results}

In this section we present a Monte Carlo study in which we examine the empirical sizes as well as the empirical powers achieved by the various tests discussed in Section 2. Section 2.1 details the simulation setting used, including the alternative distributions considered, while Section 2.2 shows the numerical results obtained together with a discussion and comparison of these results.

\subsection{Simulation setting}

We consider four different Monte Carlo settings. In the first two of these we consider the case in which only the shape parameter of the Pareto distribution requires estimation, while both the shape and scale parameters are estimated in the third and fourth settings. Furthermore, in the first and third settings we use maximum likelihood estimation in order to obtain parameter estimates, while the adjusted method of moments is used in the second and fourth settings.

We calculate empirical sizes and powers for samples of size $n=20$ and $n=30$. The empirical powers are calculated against the range of alternative distributions given in Table 1. Traditionally, these alternatives have support $(0,\infty)$. In order to ensure that the simulated data have the same support as the Pareto distribution, these alternatives are shifted by 1.

\begin{table}[!htbp!]%
\begin{center}
\caption {Summary of various choices of the alternative distributions.}
\begin{tabular}{lll}
\hline
\hline
Alternative & Density function & Notation\\
\hline
Gamma & $\frac{1}{\Gamma(\theta)}(x-1)^{\theta-1}\mathrm{e}^{-(x-1)}$ & $\Gamma(\theta)$ \\
Weibull & $\theta (x-1)^{\theta-1}\exp(-(x-1)^\theta)$ & $W(\theta)$ \\
Log-normal & $\exp\left\{-\frac{1}{2}(\log(x-1)/\theta)^2\right\}/\left\{\theta (x-1) \sqrt{2\pi}\right\}$ & $LN(\theta)$ \\
Half-normal & $\frac{\sqrt{2}}{\theta\sqrt{\pi}}\exp\left(\frac{-(x-1)^2}{2\theta^2}\right)$ & $HN(\theta)$ \\
Linear failure rate & $(1+\theta (x-1))\exp(-(x-1)-\theta (x-1)^2/2)$ & $LF(\theta)$ \\
Beta exponential & $\theta \mathrm{e}^{-(x-1)}(1-\mathrm{e}^{-(x-1)})^{\theta-1}$ &$BE(\theta)$\\
Tilted Pareto & $\frac{1+\theta}{(x+\theta)^2}$ & $TP (\theta)$\\
Dhillon & $\frac{\theta + 1}{x+1}\exp\left\{-(\log(x+1))^{\theta+1}\right\}(\log(x+1))^\theta$ & $DH(\theta)$\\
\hline
\hline
\end{tabular}
\end{center}
\end{table}

The powers obtained against these alternative distributions are displayed in Table 3 -- 6 and 10 -- 13. For ease of reference, the entries in Table 2 below gives a brief summary of the settings used in these power tables with respect to the sample size, estimation method and the number of parameters estimated.

\begin{table}[!htbp!]%
\caption {Summary of power tables.}
\begin{centering}
\begin{tabular}{ccclcccl}
\hline
\hline
Table & $n$ & Estimation & Parameters & Table & $n$ & Estimation & Parameters\\
3 & 20 & MLE & 1 parameter & 7 & 30 & MLE & 1 parameter \\
4 & 20 & MME & 1 parameter & 8 & 30 & MME & 1 parameter \\
5 & 20 & MLE & 2 parameters & 9 & 30 & MLE & 2 parameters \\
6 & 20 & MME & 2 parameters & 10 & 30 & MME & 2 parameters \\
\hline
\hline
\end{tabular}
\par\end{centering}
\end{table}

Where MLE is used for parameter estimation, we approximate critical values using 100\,000 Monte Carlo replications. Thereafter we generate 10\,000 samples from each alternative distribution considered and we calculate the empirical powers as the percentages (rounded to the nearest integers) of these samples that resulted in the rejection of $H_0$ in (\ref{H0main}). In the case where MME is used in order to perform parameter estimation we are unable to calculate fixed critical values. As a result, we use the warp-speed bootstrap method proposed in \cite{giacomini2013warp} in order to arrive at empirical critical values and powers in this case. This technique entails the following: each Monte Carlo sample is not subject to a large number of time-consuming bootstrap replications since only one bootstrap sample is taken for each Monte Carlo replication. The warp-speed method has been used in numerous studies to evaluate the power performances of the tests, see, e.g. \cite{cockeran2021goodness} as well as \cite{ndwandwe2021new}. In this setting, we make use of 50\,000 Monte Carlo samples (which then also imply 50\,000 bootstrap replications).

A final remark regarding the numerical powers associated with the tests based on characterisation of the Pareto distribution is in order. These tests, see Section 2.8, are typically much more computationally expensive to evaluate than the other tests considered. As a result, it is simply not feasible to calculate numerical powers for these tests using the warp-speed bootstrap. However, note that these tests do not require parameter estimation (the test statistics are not functions of the estimated parameter values) and we simply treat these tests as if parameter estimation is performed using MLE. Consequently, we are able to, once more, compute fixed critical values. The numerical powers reported in the tables are obtained based on these fixed critical values. In order to appreciate the large difference between the computational times required for the various tests, see Table 9 in Section 5.

\subsection{Simulation results and discussion}

The discussion below is based on the results obtained in the cases where $n=20$. The remaining results, pertaining to a sample size of $n=30$, are deferred to the appendix. However, the relative performance of the tests are similar. That is, the powers associated with the latter results are higher than those associated with the cases where $n=20$, but the tests attaining high powers in a certain setting for $n=20$ also tend to perform well in the corresponding setting when $n=30$.

Before turning our attention to a general discussion of the empirical power results or a comparison between the results associated with the various settings considered, we discuss the results obtain using maximum likelihood estimation. The numerical results shown in Tables 3 and 5 indicate that each of the tests closely maintains the specified nominal significance level of 5\%. When considering the numerical powers, it is clear that the $DK_n$ test generally outperforms all of the competing tests against the majority of alternatives considered. This impressive power performance is followed closely by that of $KL_{n,10}$, which provides power close to those achieved by $DK_n$. However, it should be noted that both of the mentioned tests exhibit dismal powers against the $W(0.5)$, $LN(2.5)$ and $BE(0.5)$ distributions. The reason for this phenomenon is a matter of ongoing research. Other tests generally producing high powers include $AD_n$, $ZB_n$ and $T_n$. These tests do not seem to produce exceedingly low powers against any specific alternatives, at least not against any of the distributions included in the Monte Carlo study.

We now turn our attention to the results obtained when using the method of moments to perform parameter estimation. The results shown in Tables 4 and 6 indicate that the tests generally fail to achieve the specified nominal significance level of 5\% against the $P(1,1)$ distribution. This can be ascribed to the fact that the first moment of the $P(1,1)$ distribution does not exist. For the remaining Pareto distributions considered, all of which posses a finite first moment, the sizes of the tests closely coincide with the specified nominal significance level. The empirical powers reported in Tables 4 and 6 show that the $G_{n,2}$ test generally achieves the highest powers in the current settings. The $MA_n$ proves to be the second most powerful test against the alternatives considered, while the $S_{n,1}$, $KL_{n,10}$ and $AD_n$ tests also generally exhibits high powers against the majority of the alternatives considered. One striking feature of the reported empirical results is the noticeably poor performance of the $DK_n$ test. While this test achieved the highest power against the majority of the alternatives when employing MLEs for parameter estimation, it frequently produces the lowest power when using the MMEs. This illustrates the importance of the choice of the estimation method when testing the assumption of the Pareto distribution.

When considering the effects of sample size and the number of parameters to be estimated, the powers are influenced in the expected way. An increase in sample size results in an increase in empirical power, while the settings in which a single parameter requires estimation generally produce higher numerical powers than is the case for settings in which both parameters are estimated. When comparing the results obtained using MLE and MME, we see that one estimation method does not increase the powers associated with each of the tests. That is, changing the estimation method used from MLE to MME results in an increase in the powers of some of the tests while the other tests experience a decrease in power. The most striking example is the $DK_n$ test which shows excellent powers when using MLE while exhibiting dismal powers when using MME.


\clearpage%
\begin{landscape}

\begin{table}[!htbp] \centering 
  \caption{Numerical powers when estimating 1 parameter using MLE with $n=20$} 
  \label{tbl:1} 
  \resizebox{\columnwidth}{!}{
\begin{tabular}{@{\extracolsep{5pt}} cccccccccccccccccccc} 
\\[-1.8ex]\hline 
\hline \\[-1.8ex]
 & $KS_n$ & $CM_n$ & $AD_n$ & $MA_n$ & $ZA_n$ & $ZB_n$ & $ZC_n$ & $KL_{n,1}$ & $KL_{n,10}$ & $DK_n$ & $S_{n,0.5}$ & $S_{n,1}$ & $G_{n,0.5}$ & $G_{n,2}$ & $T_n$ & $I_{n,2}$ & $I_{n,3}$ & $I_n^{[1]}$ & $I_n^{[2]}$ \\ 
\hline \\[-1.8ex] 
$P(1,1)$     &  {5} &  {5} &  {5} &  {5} &  {5} &  {5} &  {5} &  {5} &  {5} &  {5} &  {5} &  {5} &  {5} &  {5} &  {5} &  {5} &  {5} &  {5} &  {5} \\
$P(2,1)$     &  {5} &  {5} &  {5} &  {5} &  {5} &  {5} &  {5} &  {5} &  {5} &  {5} &  {5} &  {5} &  {5} &  {5} &  {5} &  {5} &  {5} &  {5} &  {5} \\
$P(5,1)$     & 5     & 5     &  {6} & 5     & 5     & 5     & 5     & 5     & 5     & 5     &  {6} & 5     &  {6} &  {6} & 5     & 5     & 5     & 5     & 5 \\
$P(10,1)$     &  {5} &  {5} &  {5} &  {5} &  {5} &  {5} &  {5} &  {5} &  {5} &  {5} &  {5} &  {5} &  {5} &  {5} &  {5} &  {5} &  {5} &  {5} &  {5} \\
$\Gamma{(0.5)}$     & 22    & 24    & \cellcolor{black!15}\textbf{47} & 18    & 44    & \cellcolor{black!15}\textbf{50} & 3     & 17    & 1     & 1     & 25    & 23    & 44    & 23    & 25    & 30    & 40    & 44    & 6 \\
$\Gamma{(0.8)}$    & 10    & 11    & 10    & 12    & 12    & 13    & 15    & 9     & \cellcolor{black!15}\textbf{20} & \cellcolor{black!15}\textbf{20} & 10    & 10    & 7     & 9     & 12    & 8     & 7     & 7     & 13 \\
$\Gamma{(1)}$     & 25    & 30    & 25    & 31    & 29    & 28    & 35    & 17    & \cellcolor{black!15}\textbf{46} & \cellcolor{black!15}\textbf{49} & 29    & 32    & 20    & 31    & 35    & 25    & 20    & 16    & 27 \\
$\Gamma{(1.2)}$     & 45    & 56    & 51    & 56    & 55    & 54    & 58    & 30    & \cellcolor{black!15}\textbf{71} & \cellcolor{black!15}\textbf{76} & 55    & 59    & 45    & 60    & 63    & 50    & 43    & 36    & 43 \\
$W(0.5)$     & 35    & 38    & \cellcolor{black!15}\textbf{62} & 30    & 58    & \cellcolor{black!15}\textbf{63} & 4     & 26    & 1     & 1     & 39    & 36    & 59    & 36    & 36    & 43    & 56    & 58    & 9 \\
$W(0.8)$    & 9     & 9     & 10    & 11    & 10    & 12    & 14    & 8     & \cellcolor{black!15}\textbf{17} & \cellcolor{black!15}\textbf{17} & 9     & 9     & 7     & 8     & 10    & 8     & 7     & 7     & 12 \\
$W(1.2)$    & 50    & 62    & 57    & 62    & 61    & 60    & 62    & 34    & \cellcolor{black!15}\textbf{75} & \cellcolor{black!15}\textbf{80} & 61    & 65    & 50    & 65    & 69    & 56    & 48    & 40    & 46 \\
$W(1.5)$    & 82    & 92    & 90    & 92    & 92    & 91    & 89    & 66    & \cellcolor{black!15}\textbf{96} & \cellcolor{black!15}\textbf{98} & 91    & 93    & 85    & 94    & 95    & 88    & 83    & 77    & 67 \\
$LN(1)$    & 56    & 66    & 64    & 55    & \cellcolor{black!15}\textbf{80} & \cellcolor{black!15}\textbf{78} & 39    & 37    & 59    & 65    & 66    & 70    & 75    & 73    & 73    & 64    & 66    & 62    & 19 \\
$LN(1.2)$    & 26    & 30    & 27    & 23    & \cellcolor{black!15}\textbf{43} & \cellcolor{black!15}\textbf{39} & 21    & 19    & 34    & 37    & 30    & 33    & 35    & 35    & 37    & 29    & 28    & 27    & 11 \\
$LN(1.5)$    & 7     & 7     & 6     & 7     & 10    & 9     & 9     & 9     & \cellcolor{black!15}\textbf{13} & \cellcolor{black!15}\textbf{13} & 8     & 8     & 7     & 8     & 9     & 7     & 6     & 6     & 7 \\
$LN(2.5)$    & 27    & 30    & \cellcolor{black!15}\textbf{43} & 24    & 28    & 31    & 3     & 14    & 1     & 0     & 30    & 28    & \cellcolor{black!15}\textbf{45} & 30    & 27    & 31    & 42    & 41    & 10 \\
$LFR(0.2)$    & 32    & 39    & 34    & 41    & 36    & 35    & 47    & 22    & \cellcolor{black!15}\textbf{58} & \cellcolor{black!15}\textbf{62} & 38    & 40    & 26    & 39    & 44    & 32    & 26    & 20    & 38 \\
$LFR(0.5)$    & 39    & 48    & 42    & 50    & 45    & 44    & 55    & 27    & \cellcolor{black!15}\textbf{67} & \cellcolor{black!15}\textbf{70} & 47    & 49    & 32    & 48    & 53    & 40    & 32    & 25    & 46 \\
$LFR(0.8)$    & 43    & 53    & 47    & 55    & 50    & 48    & 61    & 29    & \cellcolor{black!15}\textbf{72} & \cellcolor{black!15}\textbf{76} & 52    & 55    & 36    & 53    & 59    & 46    & 36    & 29    & 50 \\
$LFR(1)$    & 44    & 55    & 49    & 57    & 51    & 50    & 62    & 31    & \cellcolor{black!15}\textbf{73} & \cellcolor{black!15}\textbf{77} & 53    & 56    & 38    & 55    & 61    & 47    & 38    & 31    & 51 \\
$BE(0.5)$    & 19    & 20    & \cellcolor{black!15}\textbf{42} & 16    & 40    & \cellcolor{black!15}\textbf{47} & 3     & 17    & 2     & 2     & 22    & 18    & 39    & 18    & 19    & 25    & 35    & 39    & 5 \\
$BE(0.8)$    & 11    & 12    & 11    & 14    & 12    & 14    & 17    & 10    & \cellcolor{black!15}\textbf{24} & \cellcolor{black!15}\textbf{24} & 12    & 11    & 7     & 10    & 13    & 9     & 7     & 7     & 15 \\
$BE(1)$    & 24    & 30    & 25    & 31    & 28    & 27    & 35    & 17    & \cellcolor{black!15}\textbf{46} & \cellcolor{black!15}\textbf{49} & 29    & 31    & 20    & 30    & 35    & 25    & 19    & 16    & 28 \\
$BE(1.5)$    & 67    & 79    & 76    & 78    & 80    & 80    & 73    & 48    & \cellcolor{black!15}\textbf{87} & \cellcolor{black!15}\textbf{91} & 79    & 83    & 72    & 84    & 85    & 75    & 69    & 62    & 51 \\
$TP(0.5)$    & 7     & 8     & 6     & 7     & 8     & 8     & 8     & 7     & \cellcolor{black!15}\textbf{11} & \cellcolor{black!15}\textbf{11} & 8     & 7     & 6     & 7     & 9     & 7     & 6     & 6     & 6 \\
$TP(1,1)$    & 12    & 13    & 10    & 12    & 14    & 12    & 12    & 8     & \cellcolor{black!15}\textbf{18} & \cellcolor{black!15}\textbf{19} & 13    & 13    & 10    & 13    & 16    & 11    & 10    & 9     & 10 \\
$TP(2,1)$    & 22    & 26    & 22    & 23    & 25    & 24    & 21    & 12    & \cellcolor{black!15}\textbf{30} & \cellcolor{black!15}\textbf{34} & 26    & 27    & 20    & 26    & \cellcolor{black!15}\textbf{30} & 23    & 20    & 18    & 16 \\
$TP(3)$    & 32    & 37    & 32    & 34    & 35    & 34    & 29    & 17    & 41    & \cellcolor{black!15}\textbf{47} & 37    & 39    & 30    & 38    & \cellcolor{black!15}\textbf{43} & 33    & 29    & 26    & 22 \\
$D(0.2)$    & 12    & 13    & 10    & 11    & 16    & 14    & 11    & 9     & \cellcolor{black!15}\textbf{17} & \cellcolor{black!15}\textbf{19} & 13    & 14    & 12    & 14    & 16    & 12    & 10    & 10    & 8 \\
$D(0.4)$    & 30    & 35    & 31    & 30    & 40    & 37    & 26    & 18    & 39    & \cellcolor{black!15}\textbf{44} & 35    & 37    & 34    & 39    & \cellcolor{black!15}\textbf{41} & 32    & 31    & 28    & 17 \\
$D(0.6)$    & 52    & 62    & 58    & 54    & 66    & 65    & 44    & 32    & 62    & \cellcolor{black!15}\textbf{69} & 62    & 66    & 60    & 67    & \cellcolor{black!15}\textbf{68} & 58    & 56    & 51    & 26 \\
$D(0.8)$    & 70    & 81    & 79    & 76    & 84    & 83    & 63    & 48    & 80    & \cellcolor{black!15}\textbf{87} & 81    & 85    & 79    & \cellcolor{black!15}\textbf{86} & \cellcolor{black!15}\textbf{86} & 78    & 75    & 70    & 37 \\
$HN(0.8)$    & 49    & 59    & 54    & 62    & 55    & 54    & 67    & 34    & \cellcolor{black!15}\textbf{77} & \cellcolor{black!15}\textbf{80} & 58    & 61    & 41    & 59    & 65    & 52    & 41    & 33    & 56 \\
$HN(1)$    & 54    & 65    & 59    & 68    & 60    & 59    & 73    & 39    & \cellcolor{black!15}\textbf{83} & \cellcolor{black!15}\textbf{85} & 63    & 66    & 45    & 64    & 70    & 57    & 45    & 37    & 62 \\
\hline \\[-1.8ex] 
\end{tabular}
}
\end{table}

\begin{table}[!htbp] \centering 
    \caption{Numerical powers when estimating 1 parameter using MME with $n=20$} 
  \label{tbl:2} 
  \resizebox{\columnwidth}{!}{
\begin{tabular}{@{\extracolsep{5pt}} cccccccccccccccccccc} 
\\[-1.8ex]\hline 
\hline \\[-1.8ex]
 & $KS_n$ & $CM_n$ & $AD_n$ & $MA_n$ & $ZA_n$ & $ZB_n$ & $ZC_n$ & $KL_{n,1}$ & $KL_{n,10}$ & $DK_n$ & $S_{n,0.5}$ & $S_{n,1}$ & $G_{n,0.5}$ & $G_{n,2}$ & $T_n$ & $I_{n,2}$ & $I_{n,3}$ & $I_n^{[1]}$ & $I_n^{[2]}$ \\ 
\hline \\[-1.8ex] 
$P(1,1)$     & 9     & 10    & 12    &  {15} & 10    & 11    &  {14} & 6     & 4     & 1     & 10    & 11    & 6     & 11    & 5     & 5     & 5     & 5     & 5 \\
$P(2,1)$     &  {5} &  {5} &  {5} &  {5} &  {5} &  {5} & 4     &  {5} &  {5} & 2     &  {5} &  {5} &  {5} &  {5} &  {5} &  {5} &  {5} &  {5} &  {5} \\
$P(5,1)$     &  {5} &  {5} &  {5} &  {5} &  {5} &  {5} &  {5} &  {5} &  {5} & 4     &  {5} &  {5} &  {5} &  {5} &  {5} &  {5} &  {5} &  {5} &  {5} \\
$P(10,1)$     &  {5} &  {5} &  {5} &  {5} &  {5} &  {5} &  {5} &  {5} &  {5} &  {5} &  {5} &  {5} &  {5} & 4     &  {5} &  {5} &  {5} &  {5} &  {5} \\
$\Gamma{(0.5)}$     & 15    & 15    & 37    & 11    & 38    & \cellcolor{black!15}\textbf{47} & 2     & 17    & 1     & 0     & 17    & 13    & 16    & 5     & 25    & 30    & 40    & \cellcolor{black!15}\textbf{44} & 6 \\
$\Gamma{(0.8)}$     & 15    & 17    & 16    & \cellcolor{black!15}\textbf{20} & 11    & 14    & 9     & 9     & 17    & 3     & 16    & 17    & 12    & \cellcolor{black!15}\textbf{20} & 12    & 8     & 7     & 7     & 13 \\
$\Gamma{(1)}$     & 38    & 44    & 42    & \cellcolor{black!15}\textbf{47} & 33    & 35    & 17    & 18    & 41    & 8     & 43    & 45    & 37    & \cellcolor{black!15}\textbf{51} & 35    & 25    & 20    & 16    & 27 \\
$\Gamma{(1.2)}$     & 65    & 74    & 73    & 75    & 64    & 66    & 26    & 35    & 67    & 16    & 72    & \cellcolor{black!15}\textbf{76} & 69    & \cellcolor{black!15}\textbf{80} & 63    & 50    & 43    & 36    & 43 \\
$W(0.5)$     & 16    & 15    & 43    & 11    & 45    & \cellcolor{black!15}\textbf{56} & 2     & 26    & 0     & 0     & 19    & 11    & 26    & 6     & 36    & 43    & \cellcolor{black!15}\textbf{56} & \cellcolor{black!15}\textbf{58} & 9 \\
$W(0.8)$    & 15    & 17    & 17    & \cellcolor{black!15}\textbf{21} & 11    & 14    & 5     & 9     & 13    & 1     & 16    & 17    & 10    & \cellcolor{black!15}\textbf{19} & 10    & 8     & 7     & 7     & 12 \\
$W(1.2)$    & 66    & 75    & 73    & 76    & 65    & 68    & 38    & 37    & 73    & 29    & 73    & \cellcolor{black!15}\textbf{77} & 72    & \cellcolor{black!15}\textbf{80} & 69    & 56    & 48    & 40    & 46 \\
$W(1.5)$    & 91    & \cellcolor{black!15}\textbf{96} & \cellcolor{black!15}\textbf{96} & \cellcolor{black!15}\textbf{96} & 93    & 94    & 70    & 68    & \cellcolor{black!15}\textbf{96} & 73    & 95    & \cellcolor{black!15}\textbf{96} & \cellcolor{black!15}\textbf{96} & \cellcolor{black!15}\textbf{97} & 95    & 88    & 83    & 77    & 67 \\
$LN(1)$    & 73    & 82    & 83    & 75    & \cellcolor{black!15}\textbf{90} & \cellcolor{black!15}\textbf{87} & 10    & 43    & 54    & 13    & 80    & 83    & 85    & 85    & 73    & 64    & 66    & 62    & 19 \\
$LN(1.2)$    & 42    & 51    & 52    & 47    & \cellcolor{black!15}\textbf{59} & 55    & 3     & 23    & 28    & 4     & 48    & 53    & 51    & \cellcolor{black!15}\textbf{57} & 37    & 29    & 28    & 27    & 11 \\
$LN(1.5)$    & 17    & 21    & 22    & \cellcolor{black!15}\textbf{25} & 21    & 19    & 2     & 11    & 9     & 1     & 19    & 22    & 15    & \cellcolor{black!15}\textbf{25} & 9     & 7     & 6     & 6     & 7 \\
$LN(2.5)$    & 11    & 9     & 26    & 26    & 22    & 34    & 36    & 19    & 0     & 0     & 12    & 7     & 9     & 5     & 27    & 31    & \cellcolor{black!15}\textbf{42} & \cellcolor{black!15}\textbf{41} & 10 \\
$LFR(0.2)$    & 45    & 53    & 50    & \cellcolor{black!15}\textbf{57} & 40    & 43    & 28    & 23    & 55    & 15    & 52    & 55    & 47    & \cellcolor{black!15}\textbf{59} & 44    & 32    & 26    & 20    & 38 \\
$LFR(0.5)$    & 51    & 59    & 56    & 63    & 47    & 49    & 39    & 28    & \cellcolor{black!15}\textbf{65} & 25    & 58    & 60    & 54    & \cellcolor{black!15}\textbf{65} & 53    & 40    & 32    & 25    & 46 \\
$LFR(0.8)$    & 54    & 63    & 59    & 66    & 50    & 53    & 45    & 30    & \cellcolor{black!15}\textbf{70} & 32    & 61    & 64    & 59    & \cellcolor{black!15}\textbf{68} & 59    & 46    & 36    & 29    & 50 \\
$LFR(1)$    & 57    & 66    & 63    & 69    & 53    & 56    & 49    & 32    & \cellcolor{black!15}\textbf{73} & 38    & 64    & 67    & 63    & \cellcolor{black!15}\textbf{70} & 61    & 47    & 38    & 31    & 51 \\
$BE(0.5)$    & 12    & 11    & 32    & 10    & 34    & \cellcolor{black!15}\textbf{44} & 2     & 17    & 1     & 0     & 13    & 9     & 12    & 4     & 19    & 25    & 35    & \cellcolor{black!15}\textbf{39} & 5 \\
$BE(0.8)$    & 17    & 19    & 18    & \cellcolor{black!15}\textbf{24} & 13    & 16    & 10    & 10    & 19    & 3     & 19    & 20    & 14    & \cellcolor{black!15}\textbf{23} & 13    & 9     & 7     & 7     & 15 \\
$BE(1)$    & 38    & 45    & 42    & \cellcolor{black!15}\textbf{48} & 34    & 35    & 17    & 19    & 41    & 8     & 43    & 46    & 38    & \cellcolor{black!15}\textbf{51} & 35    & 25    & 19    & 16    & 28 \\
$BE(1.5)$    & 83    & 91    & 91    & 90    & 86    & 88    & 35    & 53    & 84    & 30    & 90    & \cellcolor{black!15}\textbf{92} & 90    & \cellcolor{black!15}\textbf{94} & 85    & 75    & 69    & 62    & 51 \\
$TP(0.5)$    & 27    & 31    & \cellcolor{black!15}\textbf{35} & \cellcolor{black!15}\textbf{39} & 30    & 33    & 19    & 13    & 7     & 0     & 30    & 32    & 16    & 32    & 9     & 7     & 6     & 6     & 6 \\
$TP(1,1)$    & 46    & 53    & \cellcolor{black!15}\textbf{58} & \cellcolor{black!15}\textbf{62} & 52    & 56    & 25    & 22    & 11    & 0     & 51    & 54    & 29    & 54    & 16    & 11    & 10    & 9     & 10 \\
$TP(2,1)$    & 76    & 82    & \cellcolor{black!15}\textbf{86} & \cellcolor{black!15}\textbf{88} & 81    & 84    & 38    & 44    & 19    & 0     & 80    & 82    & 55    & 82    & 30    & 23    & 20    & 18    & 16 \\
$TP(3)$    & 90    & 94    & \cellcolor{black!15}\textbf{96} & \cellcolor{black!15}\textbf{97} & 94    & 95    & 50    & 64    & 28    & 0     & 93    & 94    & 72    & 94    & 43    & 33    & 29    & 26    & 22 \\
$D(0.2)$    & 26    & 31    & \cellcolor{black!15}\textbf{32} & \cellcolor{black!15}\textbf{32} & 29    & 27    & 4     & 12    & 13    & 2     & 30    & \cellcolor{black!15}\textbf{32} & 24    & \cellcolor{black!15}\textbf{34} & 16    & 12    & 10    & 10    & 8 \\
$D(0.4)$    & 50    & 59    & 59    & 57    & 57    & 55    & 3     & 23    & 31    & 4     & 57    & \cellcolor{black!15}\textbf{61} & 53    & \cellcolor{black!15}\textbf{64} & 41    & 32    & 31    & 28    & 17 \\
$D(0.6)$    & 73    & 82    & 82    & 79    & 80    & 79    & 8     & 40    & 55    & 9     & 80    & \cellcolor{black!15}\textbf{83} & 79    & \cellcolor{black!15}\textbf{86} & 68    & 58    & 56    & 51    & 26 \\
$D(0.8)$    & 89    & 94    & 94    & 93    & 93    & 93    & 16    & 58    & 75    & 17    & 93    & \cellcolor{black!15}\textbf{95} & 93    & \cellcolor{black!15}\textbf{96} & 86    & 78    & 75    & 70    & 37 \\
$HN(0.8)$    & 59    & 68    & 65    & \cellcolor{black!15}\textbf{72} & 55    & 58    & 55    & 35    & \cellcolor{black!15}\textbf{76} & 42    & 66    & 69    & 65    & \cellcolor{black!15}\textbf{72} & 65    & 52    & 41    & 33    & 56 \\
$HN(1)$    & 67    & 76    & 74    & 80    & 63    & 66    & 57    & 40    & \cellcolor{black!15}\textbf{82} & 40    & 75    & 77    & 71    & \cellcolor{black!15}\textbf{81} & 70    & 57    & 45    & 37    & 62 \\
\hline \\[-1.8ex] 
\end{tabular} 
}
\end{table}

\begin{table}[!htbp] \centering 
  \caption{Numerical powers when estimating 2 parameter using MLE with $n=20$} 
  \label{tbl:3} 
  \resizebox{\columnwidth}{!}{
\begin{tabular}{@{\extracolsep{5pt}} cccccccccccccccccccc} 
\\[-1.8ex]\hline 
\hline \\[-1.8ex]
 & $KS_n$ & $CM_n$ & $AD_n$ & $MA_n$ & $ZA_n$ & $ZB_n$ & $ZC_n$ & $KL_{n,1}$ & $KL_{n,10}$ & $DK_n$ & $S_{n,0.5}$ & $S_{n,1}$ & $G_{n,0.5}$ & $G_{n,2}$ & $T_n$ & $I_{n,2}$ & $I_{n,3}$ & $I_n^{[1]}$ & $I_n^{[2]}$ \\ 
\hline \\[-1.8ex] 
$P(1,1)$     &  {5} &  {5} &  {5} &  {5} &  {5} &  {5} &  {5} &  {5} &  {5} &  {5} &  {5} &  {5} &  {5} &  {5} &  {5} &  {5} &  {5} &  {5} &  {5} \\
$P(2,1)$     &  {5} &  {5} &  {5} &  {5} &  {5} &  {5} &  {5} &  {5} &  {5} &  {5} &  {5} &  {5} &  {5} &  {5} &  {5} &  {5} &  {5} &  {5} &  {5} \\
$P(5,1)$     &  {5} &  {5} &  {5} &  {5} &  {5} &  {5} &  {5} &  {5} &  {5} &  {5} &  {5} &  {5} &  {5} &  {5} &  {5} &  {5} &  {5} &  {5} &  {5} \\
$P(10,1)$     &  {5} &  {5} &  {5} &  {5} &  {5} &  {5} &  {5} &  {5} &  {5} &  {5} &  {5} &  {5} &  {5} &  {5} &  {5} &  {5} &  {5} &  {5} &  {5} \\
$\Gamma{(0.5)}$     & 23    & 25    & \cellcolor{black!15}\textbf{39} & 19    & \cellcolor{black!15}\textbf{38} & 37    & 2     & 26    & 2     & 2     & 25    & 23    & 37    & 23    & 26    & 27    & 33    & 28    & 9 \\
$\Gamma{(0.8)}$     & 7     & 7     & 5     & 8     & 7     & 5     & 19    & 10    & \cellcolor{black!15}\textbf{21} & \cellcolor{black!15}\textbf{22} & 7     & 5     & 2     & 2     & 6     & 3     & 2     & 7     & 13 \\
$\Gamma{(1)}$     & 15    & 15    & 5     & 18    & 9     & 3     & 38    & 13    & \cellcolor{black!15}\textbf{41} & \cellcolor{black!15}\textbf{44} & 15    & 13    & 0     & 7     & 15    & 7     & 3     & 14    & 26 \\
$\Gamma{(1.2)}$     & 26    & 29    & 11    & 33    & 18    & 4     & 58    & 18    & \cellcolor{black!15}\textbf{60} & \cellcolor{black!15}\textbf{64} & 28    & 26    & 0     & 16    & 30    & 16    & 7     & 26    & 41 \\
$W(0.5)$     & 35    & 38    & \cellcolor{black!15}\textbf{54} & 30    & \cellcolor{black!15}\textbf{52} & 51    & 3     & 36    & 2     & 1     & 39    & 36    & \cellcolor{black!15}\textbf{52} & 35    & 37    & 39    & 47    & 41    & 13 \\
$W(0.8)$    & 6     & 6     & 5     & 8     & 7     & 6     & 17    & 10    & \cellcolor{black!15}\textbf{19} & \cellcolor{black!15}\textbf{20} & 7     & 4     & 3     & 2     & 6     & 3     & 3     & 7     & 12 \\
$W(1.2)$    & 28    & 32    & 13    & 36    & 20    & 4     & 61    & 19    & \cellcolor{black!15}\textbf{63} & \cellcolor{black!15}\textbf{68} & 32    & 29    & 0     & 18    & 33    & 18    & 8     & 29    & 44 \\
$W(1.5)$    & 48    & 56    & 30    & 60    & 41    & 11    & \cellcolor{black!15}\textbf{82} & 32    & \cellcolor{black!15}\textbf{82} & \cellcolor{black!15}\textbf{87} & 55    & 53    & 1     & 39    & 58    & 38    & 19    & 50    & 66 \\
$LN(1)$    & 14    & 14    & 4     & 14    & 7     & 2     & 28    & 8     & \cellcolor{black!15}\textbf{30} & \cellcolor{black!15}\textbf{34} & 14    & 12    & 0     & 7     & 15    & 8     & 4     & 20    & 16 \\
$LN(1.2)$    & 7     & 7     & 2     & 7     & 4     & 2     & 17    & 6     & \cellcolor{black!15}\textbf{20} & \cellcolor{black!15}\textbf{21} & 7     & 6     & 0     & 3     & 7     & 3     & 2     & 11    & 10 \\
$LN(1.5)$    & 4     & 4     & 3     & 4     & 3     & 3     & 8     & 6     & \cellcolor{black!15}\textbf{10} & \cellcolor{black!15}\textbf{10} & 4     & 3     & 2     & 2     & 4     & 3     & 2     & 5     & 6 \\
$LN(2.5)$    & 29    & 32    & \cellcolor{black!15}\textbf{43} & 26    & 38    & 34    & 2     & 21    & 1     & 1     & 33    & 31    & \cellcolor{black!15}\textbf{43} & 32    & 32    & 33    & 39    & 30    & 13 \\
$LFR(0.2)$    & 20    & 22    & 9     & 26    & 14    & 4     & 50    & 17    & \cellcolor{black!15}\textbf{53} & \cellcolor{black!15}\textbf{57} & 21    & 18    & 0     & 11    & 22    & 11    & 4     & 19    & 37 \\
$LFR(0.5)$    & 25    & 28    & 11    & 33    & 18    & 5     & 59    & 19    & \cellcolor{black!15}\textbf{61} & \cellcolor{black!15}\textbf{64} & 27    & 24    & 0     & 14    & 28    & 15    & 6     & 23    & 44 \\
$LFR(0.8)$    & 28    & 32    & 13    & 37    & 21    & 5     & 64    & 21    & \cellcolor{black!15}\textbf{66} & \cellcolor{black!15}\textbf{69} & 31    & 28    & 0     & 17    & 32    & 17    & 7     & 26    & 48 \\
$LFR(1)$    & 29    & 33    & 14    & 38    & 22    & 6     & 64    & 22    & \cellcolor{black!15}\textbf{67} & \cellcolor{black!15}\textbf{71} & 32    & 29    & 0     & 18    & 33    & 18    & 8     & 27    & 49 \\
$BE(0.5)$    & 20    & 21    & \cellcolor{black!15}\textbf{35} & 16    & \cellcolor{black!15}\textbf{34} & \cellcolor{black!15}\textbf{34} & 3     & 26    & 4     & 3     & 22    & 18    & 32    & 17    & 20    & 22    & 28    & 24    & 7 \\
$BE(0.8)$    & 8     & 8     & 4     & 10    & 6     & 5     & 22    & 11    & \cellcolor{black!15}\textbf{25} & \cellcolor{black!15}\textbf{25} & 7     & 5     & 2     & 2     & 6     & 3     & 2     & 7     & 15 \\
$BE(1)$    & 15    & 15    & 6     & 19    & 9     & 3     & 39    & 13    & \cellcolor{black!15}\textbf{42} & \cellcolor{black!15}\textbf{45} & 15    & 13    & 0     & 7     & 15    & 8     & 3     & 14    & 27 \\
$BE(1.5)$    & 34    & 40    & 17    & 43    & 25    & 6     & 68    & 22    & \cellcolor{black!15}\textbf{69} & \cellcolor{black!15}\textbf{74} & 39    & 37    & 0     & 25    & 41    & 24    & 11    & 37    & 49 \\
$TP(0.5)$    & 5     & 4     & 3     & 5     & 3     & 4     & 8     & 6     & \cellcolor{black!15}\textbf{9} & \cellcolor{black!15}\textbf{10} & 4     & 3     & 2     & 2     & 4     & 3     & 2     & 6     & 6 \\
$TP(1,1)$    & 7     & 6     & 2     & 7     & 4     & 3     & 15    & 6     & \cellcolor{black!15}\textbf{16} & \cellcolor{black!15}\textbf{17} & 6     & 5     & 1     & 3     & 6     & 3     & 2     & 8     & 9 \\
$TP(2,1)$    & 11    & 11    & 4     & 12    & 6     & 2     & 24    & 8     & \cellcolor{black!15}\textbf{25} & \cellcolor{black!15}\textbf{29} & 11    & 9     & 0     & 5     & 12    & 6     & 2     & 15    & 15 \\
$TP(3)$    & 17    & 18    & 6     & 18    & 9     & 3     & \cellcolor{black!15}\textbf{34} & 11    & \cellcolor{black!15}\textbf{34} & \cellcolor{black!15}\textbf{39} & 17    & 15    & 0     & 8     & 18    & 10    & 4     & 21    & 21 \\
$D(0.2)$    & 5     & 5     & 2     & 5     & 3     & 3     & 12    & 6     & \cellcolor{black!15}\textbf{14} & \cellcolor{black!15}\textbf{15} & 5     & 4     & 1     & 2     & 5     & 3     & 2     & 8     & 8 \\
$D(0.4)$    & 12    & 12    & 4     & 12    & 6     & 2     & 26    & 8     & \cellcolor{black!15}\textbf{28} & \cellcolor{black!15}\textbf{31} & 12    & 10    & 0     & 6     & 13    & 7     & 3     & 17    & 16 \\
$D(0.6)$    & 20    & 22    & 8     & 21    & 12    & 3     & 40    & 11    & \cellcolor{black!15}\textbf{41} & \cellcolor{black!15}\textbf{47} & 22    & 20    & 0     & 12    & 23    & 13    & 6     & 27    & 24 \\
$D(0.8)$    & 29    & 33    & 14    & 33    & 19    & 5     & \cellcolor{black!15}\textbf{55} & 16    & 54    & \cellcolor{black!15}\textbf{62} & 32    & 30    & 0     & 21    & 34    & 20    & 11    & 37    & 35 \\
$HN(0.8)$    & 32    & 37    & 17    & 43    & 26    & 6     & 69    & 24    & \cellcolor{black!15}\textbf{71} & \cellcolor{black!15}\textbf{74} & 36    & 32    & 0     & 21    & 37    & 21    & 8     & 29    & 54 \\
$HN(1)$    & 36    & 42    & 19    & 48    & 29    & 8     & 75    & 27    & \cellcolor{black!15}\textbf{76} & \cellcolor{black!15}\textbf{79} & 40    & 37    & 0     & 24    & 42    & 24    & 10    & 32    & 60 \\
\hline \\[-1.8ex] 
\end{tabular} 
}
\end{table}

\begin{table}[!htbp] \centering 
  \caption{Numerical powers when estimating 2 parameter using MME with $n=20$}   \label{tbl:4} 
  \resizebox{\columnwidth}{!}{
\begin{tabular}{@{\extracolsep{5pt}} cccccccccccccccccccc} 
\\[-1.8ex]\hline 
\hline \\[-1.8ex]
 & $KS_n$ & $CM_n$ & $AD_n$ & $MA_n$ & $ZA_n$ & $ZB_n$ & $ZC_n$ & $KL_{n,1}$ & $KL_{n,10}$ & $DK_n$ & $S_{n,0.5}$ & $S_{n,1}$ & $G_{n,0.5}$ & $G_{n,2}$ & $T_n$ & $I_{n,2}$ & $I_{n,3}$ & $I_n^{[1]}$ & $I_n^{[2]}$ \\ 
\hline \\[-1.8ex] 
$P(1,1)$     & 9     & 11    & 12    &  {15} & 13    &  {14} &  {14} & 6     & 3     & 1     & 10    & 11    & 6     & 11    & 5     & 5     & 5     & 5     & 5 \\
$P(2,1)$     &  {5} &  {5} &  {5} &  {5} &  {5} &  {5} & 4     &  {5} &  {5} & 2     &  {5} &  {5} &  {5} &  {5} &  {5} &  {5} &  {5} &  {5} &  {5} \\
$P(5,1)$     &  {5} &  {5} &  {5} &  {5} &  {5} &  {5} &  {5} &  {5} &  {5} & 4     &  {5} &  {5} &  {5} &  {5} &  {5} &  {5} &  {5} &  {5} &  {5} \\
$P(10,1)$     &  {5} &  {5} &  {5} &  {5} &  {5} &  {5} &  {5} &  {5} &  {5} & 4     &  {5} &  {5} &  {5} & 4     &  {5} &  {5} &  {5} &  {5} &  {5} \\
$\Gamma{(0.5)}$     & 10    & 10    & 15    & 8     & 10    & 11    & 2     & 26    & 2     & 0     & 10    & 8     & 9     & 4     & 26    & 27    & \cellcolor{black!15}\textbf{33} & \cellcolor{black!15}\textbf{28} & 9 \\
$\Gamma{(0.8)}$     & 16    & 18    & 17    & \cellcolor{black!15}\textbf{21} & 15    & 18    & 9     & 10    & 18    & 3     & 17    & 19    & 15    & \cellcolor{black!15}\textbf{23} & 6     & 3     & 2     & 7     & 13 \\
$\Gamma{(1)}$     & 34    & 40    & 39    & \cellcolor{black!15}\textbf{43} & 36    & 40    & 15    & 13    & 36    & 7     & 39    & 42    & 36    & \cellcolor{black!15}\textbf{47} & 15    & 7     & 3     & 14    & 26 \\
$\Gamma{(1.2)}$     & 53    & 61    & 61    & \cellcolor{black!15}\textbf{63} & 58    & 62    & 22    & 19    & 55    & 12    & 60    & \cellcolor{black!15}\textbf{63} & 56    & \cellcolor{black!15}\textbf{68} & 30    & 16    & 7     & 26    & 41 \\
$W(0.5)$     & 12    & 10    & 18    & 10    & 14    & 16    & 1     & 37    & 1     & 0     & 12    & 8     & 12    & 5     & 37    & 39    & \cellcolor{black!15}\textbf{47} & \cellcolor{black!15}\textbf{41} & 13 \\
$W(0.8)$    & 17    & 19    & 19    & \cellcolor{black!15}\textbf{23} & 15    & 18    & 5     & 10    & 14    & 1     & 18    & 19    & 13    & \cellcolor{black!15}\textbf{22} & 6     & 3     & 3     & 7     & 12 \\
$W(1.2)$    & 53    & 61    & 61    & 63    & 58    & 63    & 31    & 19    & 59    & 20    & 60    & \cellcolor{black!15}\textbf{64} & 60    & \cellcolor{black!15}\textbf{68} & 33    & 18    & 8     & 29    & 44 \\
$W(1.5)$    & 71    & 79    & 79    & 80    & 78    & \cellcolor{black!15}\textbf{82} & 54    & 32    & 81    & 46    & 78    & 80    & 79    & \cellcolor{black!15}\textbf{82} & 58    & 38    & 19    & 50    & 66 \\
$LN(1)$    & 33    & 40    & 39    & 38    & 38    & 39    & 7     & 9     & 25    & 7     & 39    & \cellcolor{black!15}\textbf{42} & 38    & \cellcolor{black!15}\textbf{45} & 15    & 8     & 4     & 20    & 16 \\
$LN(1.2)$    & 23    & 28    & 28    & 28    & 27    & 27    & 3     & 7     & 15    & 3     & 26    & \cellcolor{black!15}\textbf{29} & 24    & \cellcolor{black!15}\textbf{32} & 7     & 3     & 2     & 11    & 10 \\
$LN(1.5)$    & 14    & 16    & 18    & \cellcolor{black!15}\textbf{21} & 16    & 16    & 2     & 6     & 6     & 1     & 16    & 17    & 10    & \cellcolor{black!15}\textbf{19} & 4     & 3     & 2     & 5     & 6 \\
$LN(2.5)$    & 12    & 11    & 24    & 30    & 31    & \cellcolor{black!15}\textbf{37} & \cellcolor{black!15}\textbf{37} & 28    & 1     & 0     & 12    & 9     & 4     & 8     & 32    & 33    & \cellcolor{black!15}\textbf{39} & 30    & 13 \\
$LFR(0.2)$    & 41    & 48    & 48    & \cellcolor{black!15}\textbf{51} & 45    & 50    & 25    & 16    & 49    & 12    & 47    & 50    & 45    & \cellcolor{black!15}\textbf{56} & 22    & 11    & 4     & 19    & 37 \\
$LFR(0.5)$    & 46    & 54    & 53    & 57    & 51    & 56    & 35    & 19    & \cellcolor{black!15}\textbf{58} & 19    & 52    & 55    & 51    & \cellcolor{black!15}\textbf{60} & 28    & 15    & 6     & 23    & 44 \\
$LFR(0.8)$    & 48    & 57    & 56    & 60    & 54    & 60    & 41    & 20    & \cellcolor{black!15}\textbf{63} & 25    & 55    & 59    & 56    & \cellcolor{black!15}\textbf{63} & 32    & 17    & 7     & 26    & 48 \\
$LFR(1)$    & 51    & 59    & 58    & 62    & 58    & 63    & 44    & 22    & \cellcolor{black!15}\textbf{66} & 29    & 58    & 61    & 58    & \cellcolor{black!15}\textbf{64} & 33    & 18    & 8     & 27    & 49 \\
$BE(0.5)$    & 8     & 8     & 13    & 8     & 8     & 10    & 3     & \cellcolor{black!15}\textbf{25} & 3     & 0     & 9     & 6     & 7     & 4     & 20    & 22    & \cellcolor{black!15}\textbf{28} & 24    & 7 \\
$BE(0.8)$    & 18    & 21    & 21    & \cellcolor{black!15}\textbf{25} & 18    & 21    & 10    & 11    & 21    & 3     & 20    & 22    & 17    & \cellcolor{black!15}\textbf{26} & 6     & 3     & 2     & 7     & 15 \\
$BE(1)$    & 35    & 41    & 40    & \cellcolor{black!15}\textbf{43} & 36    & 41    & 16    & 13    & 37    & 6     & 39    & 42    & 36    & \cellcolor{black!15}\textbf{47} & 15    & 8     & 3     & 14    & 27 \\
$BE(1.5)$    & 63    & 72    & 72    & 72    & 69    & 72    & 28    & 23    & 64    & 20    & 71    & \cellcolor{black!15}\textbf{74} & 69    & \cellcolor{black!15}\textbf{77} & 41    & 24    & 11    & 37    & 49 \\
$TP(0.5)$    & 25    & 29    & 33    & \cellcolor{black!15}\textbf{37} & \cellcolor{black!15}\textbf{36} & \cellcolor{black!15}\textbf{36} & 19    & 10    & 6     & 0     & 28    & 30    & 15    & 30    & 4     & 3     & 2     & 6     & 6 \\
$TP(1,1)$    & 42    & 48    & 54    & \cellcolor{black!15}\textbf{57} & 56    & \cellcolor{black!15}\textbf{57} & 24    & 16    & 9     & 0     & 46    & 49    & 25    & 49    & 6     & 3     & 2     & 8     & 9 \\
$TP(2,1)$    & 68    & 75    & 80    & \cellcolor{black!15}\textbf{82} & \cellcolor{black!15}\textbf{82} & \cellcolor{black!15}\textbf{82} & 34    & 32    & 15    & 0     & 74    & 76    & 47    & 75    & 12    & 6     & 2     & 15    & 15 \\
$TP(3)$    & 82    & 88    & 91    & \cellcolor{black!15}\textbf{92} & \cellcolor{black!15}\textbf{92} & \cellcolor{black!15}\textbf{92} & 44    & 47    & 20    & 0     & 86    & 88    & 62    & 88    & 18    & 10    & 4     & 21    & 21 \\
$D(0.2)$    & 21    & 24    & \cellcolor{black!15}\textbf{25} & \cellcolor{black!15}\textbf{25} & 24    & 23    & 4     & 7     & 10    & 1     & 23    & \cellcolor{black!15}\textbf{25} & 18    & \cellcolor{black!15}\textbf{27} & 5     & 3     & 2     & 8     & 8 \\
$D(0.4)$    & 35    & 41    & 42    & 41    & 39    & 39    & 3     & 10    & 21    & 3     & 40    & \cellcolor{black!15}\textbf{43} & 34    & \cellcolor{black!15}\textbf{46} & 13    & 7     & 3     & 17    & 16 \\
$D(0.6)$    & 49    & 57    & 57    & 56    & 55    & 55    & 7     & 14    & 34    & 6     & 56    & \cellcolor{black!15}\textbf{59} & 51    & \cellcolor{black!15}\textbf{62} & 23    & 13    & 6     & 27    & 24 \\
$D(0.8)$    & 61    & 69    & 68    & 67    & 66    & 67    & 13    & 19    & 48    & 11    & 67    & \cellcolor{black!15}\textbf{70} & 64    & \cellcolor{black!15}\textbf{73} & 34    & 20    & 11    & 37    & 35 \\
$HN(0.8)$    & 53    & 62    & 60    & 65    & 60    & 66    & 49    & 23    & \cellcolor{black!15}\textbf{69} & 33    & 60    & 63    & 61    & \cellcolor{black!15}\textbf{67} & 37    & 21    & 8     & 29    & 54 \\
$HN(1)$    & 60    & 69    & 69    & 73    & 67    & 73    & 51    & 27    & \cellcolor{black!15}\textbf{75} & 31    & 68    & 71    & 66    & \cellcolor{black!15}\textbf{75} & 42    & 24    & 10    & 32    & 60 \\
\hline \\[-1.8ex] 
\end{tabular} 
}
\end{table}

\clearpage

\end{landscape}

\section{Practical application}

We now employ the various tests considered in order to ascertain whether or not an observed data set is compatible with the assumption of being realised from a Pareto distribution. The data set is comprised of the monetary expenses incurred as a result of wind related catastrophes in 40 separate instances during 1977, rounded to the nearest million US dollars. The data are provided in Table \ref{wind}.

\begin{table}[!htbp!]%
\caption{Wind catastrophes original data set.} \label{wind}
\begin{center}
\begin{tabular}{rrrrrrrrrr}
\hline
2 & 2 & 2 & 2 & 2 & 2 & 2 & 2 & 2 & 2 \\
2 & 2 & 3 & 3 & 3 & 3 & 4 & 4 & 4 & 5 \\
5 & 5 & 5 & 6 & 6 & 6 & 6 & 8 & 8 & 9 \\
15 & 17 & 22 & 23 & 24 & 24 & 25 & 27 & 32 & 43\\
\hline
\end{tabular}
\end{center}
\end{table}

The rounding of the recorded values in Table \ref{wind} causes unrealistic clustering in the data which may lead to problems when testing for the Pareto distribution. In order to circumvent the associated problems, we use the de-grouping algorithm discussed in \cite{allison2022distribution} as well as \cite{brazauskas2003favorable}. This algorithm replaces the values in each group of tied observations with the expected value of the order statistics of the uniform distribution with the same range. That is, if one observes $k$ identical integer values, $x$, in an interval $(l=x-1/2,u=x+1/2)$, we replace these values by
\begin{equation*}
    \left(\frac{k+1-j}{k+1}\right)l+\left(\frac{j}{k+1}\right)u,
\end{equation*}
for $j \in \{1,\dots,k\}$. We emphasise that this de-grouping algorithm does not change the mean of the data set. The de-grouped data can be found in Table \ref{winddg}.


\begin{table}[!htbp!]%
\caption{Wind catastrophes de-grouped data set.} \label{winddg}
\begin{center}
\begin{tabular}{rrrrrrrr}
\hline
1.58 & 1.65 & 1.73 & 1.81 & 1.88 & 1.96 & 2.04 & 2.12 \\
2.19 & 2.27 & 2.35 & 2.42 & 2.70 & 2.90 & 3.10 & 3.30 \\
3.75 & 4.00 & 4.25 & 4.70 & 4.90 & 5.10 & 5.30 & 5.70 \\
5.90 & 6.10 & 6.30 & 7.83 & 8.17 & 9.00 & 15.00 & 17.00 \\
22.00 & 23.00 & 23.83 & 24.17 & 25.00 & 27.00 & 32.00 & 43.00 \\
\hline
\end{tabular}
\end{center}
\end{table}

When testing the hypothesis of the Pareto distribution for the data set, we consider each of the four settings used in the Monte Carlo study presented in Section 4. That is, we test the hypothesis in both the one and two parameter cases and we use MLE as well as MME in order to arrive at parameter estimates. Note that, when fitting a one parameter distribution, the support of the distribution is assumed known. Since the observed minimum is rounded to 2, we conclude that no value less than 1.5 is possible. As a result, we fix $\sigma=1.5$ in the cases where the one parameter distribution is considered. No such assumption is necessary for the two parameter case; in this case, the value of $\sigma$ is simply estimated from the data.

When assuming that $\sigma=1.5$, the MLE of $\beta$ is calculated to be $\widehat{\beta}_n= 0.764$ while the MME is $\tilde{\beta}_n=1.194$. In the case where both $\beta$ and $\sigma$ are estimated; the MLEs are $\widehat{\beta}_n=0.796$ and $\widehat{\sigma}_n=1.053$. The corresponding MMEs are $\tilde{\beta}_n=1.202$ and $\widehat{\sigma}_n=1.031$. The empirical $p$-values associated with each of these four instances are shown in Table \ref{pvals}. These $p$-values are obtained using 10 000 samples generated using a parametric bootstrap procedure. The results associated with each of the tests considered in Section 4 are shown. The column headings used indicate the estimation method used as well as the number of parameters estimated. The final column in the table shows the time required in order to arrive at the reported $p$-values in seconds. The reported results are obtained using a 64 bit Windows 10 operating system with an Intel Core i7-7700 CPU @ 3.60 GHz with 16 GB of RAM. Note the substantial computational times associated with the tests based on characterisations of the Pareto distribution.

\begin{table}[ht] 
\centering 
  \caption{$p$-values for the wind catastrophes data set} 
  \label{pvals} 
\begin{tabular}{@{\extracolsep{5pt}} lrrrrr} 
\\[-1.8ex]\hline 
\hline
Test & MLE 1 & MME 1 & MLE 2 & MME 2 & Time \\
\hline
$KS_n$ & $0.509$ & $0.013$ & $0.547$ & $0.013$ & $2$ \\ 
$CM_n$ & $0.271$ & $0.004$ & $0.403$ & $0.004$ & $2$ \\ 
$AD_n$ & $0.242$ & $0.001$ & $0.000$ & $0.001$ & $0$ \\ 
$MA_n$ & $0.114$ & $0.000$ & $0.153$ & $0.000$ & $1$ \\ 
$ZA_n$ & $0.075$ & $0.006$ & $0.000$ & $0.002$ & $2$ \\ 
$ZB_n$ & $0.078$ & $0.002$ & $0.000$ & $0.001$ & $1$ \\ 
$ZC_n$ & $0.009$ & $0.226$ & $0.000$ & $0.229$ & $0$ \\ 
$KL_{n,1}$ & $0.395$ & $0.107$ & $0.460$ & $0.152$ & $2$ \\ 
$KL_{n,10}$ & $0.009$ & $0.009$ & $0.010$ & $0.012$ & $1$ \\ 
$DK_n$ & $0.013$ & $0.537$ & $0.014$ & $0.511$ & $8$ \\ 
$S_{n,0.5}$ & $0.299$ & $0.006$ & $0.412$ & $0.006$ & $9$ \\ 
$S_{n,1}$ & $0.171$ & $0.003$ & $0.291$ & $0.003$ & $12$ \\ 
$G_{n,0.5}$ & $0.217$ & $0.023$ & $0.604$ & $0.034$ & $22$ \\ 
$G_{n,2}$ & $0.133$ & $0.002$ & $0.278$ & $0.002$ & $33$ \\ 
$T_n$ & $0.265$ & $0.266$ & $0.616$ & $0.724$ & $183$ \\ 
$I_{n,2}$ & $0.632$ & $0.632$ & $0.917$ & $0.900$ & $311$ \\ 
$I_{n,3}$ & $0.425$ & $0.425$ & $0.912$ & $0.863$ & $6005$ \\ 
$I_n^{[1]}$ & $0.301$ & $0.303$ & $0.354$ & $0.968$ & $34662$ \\ 
$I_n^{[2]}$ & $0.051$ & $0.051$ & $0.053$ & $0.050$ & $83331$ \\ 
\hline \\[-1.8ex] 
\end{tabular} 
\end{table} 


In the interpretation of the $p$-values, we use a nominal significance level of 5\%. When assuming a known $\sigma$ of 1.5 and using MLE to estimate the value of $\beta$, the majority of the test statistics do not reject the null hypothesis. The exceptions, which reject hypothesis of the Pareto distribution, are $ZC_n$, $KL_{n,10}$, and $DK_n$. In the case where parameter estimation is performed using the MME, the situation is reversed and 11 of the 19 tests considered reject the Pareto assumption. The tests not rejecting the null hypothesis in this case are $ZC_n$, $KL_{n,1}$, $DK_n$, $T_n$, $I_{n,2}$, $I_{n,3}$, $I_n^{[1]}$ and $I_n^{[2]}$.

We now turn our attention to the case where both $\beta$ and $\sigma$ require estimation. We start by considering the results obtained using MLE. In this case the majority of the tests do not reject the null hypothesis. In fact $AD_n$, $ZA_n$, $ZB_n$, $ZC_n$, $KL_{n,10}$ and $DK_n$ reject the null hypothesis while the remaining 9 tests do not reject the null hypothesis. Finally, when considering the results associated with MME, we observe that the majority of the tests reject the hypothesis of the Pareto distribution. The exceptions to this are the $ZC_n$, $KL_{n,1}$, $DK_n$, $T_n$, $I_{n,2}$, $I_{n,3}$, $I_n^{[1]}$, and $I_n^{[2]}$ tests.

When comparing the $p$-values associated with the practical example, some further remarks are in order. The majority of the tests reject the assumption of the Pareto distribution in some settings whilst not rejecting it in others. In fact the only test which reject the null hypothesis in all four of the settings considered is $KL_{n,10}$. Interestingly, out of all of the tests for which parameter estimation is required, $KL_{n,1}$ is the only test that does not reject the hypothesis of the Pareto distribution in any of the four settings considered. However, the tests not requiring parameter estimation, that is, $T_n$, $I_{n,2}$, $I_{n,3}$, $I_n^{[1]}$ and $I_n^{[2]}$, all fail to reject the null hypothesis in each of the settings considered. In conclusion, there is evidence both for and against the assumption that the observed data are realised from a Pareto distribution.


\section{Concluding remarks}

The goal of this study is to review the existing goodness-of-fit tests for the Pareto type I distribution based on a wide range of characteristics of this distribution. Below we provide brief descriptions of these characteristics and the tests related to them. The tests based on the edf, commonly known as the traditional tests, are Kolmogov-Smirnov ($KS_n$), Cram\'{e}r-von Mises ($CV_n$), Anderson-Darling ($AD_n$) and modified Anderson-Darling ($MA_n$) tests. We also consider tests based on likelihood ratios. These tests are either weighted by some function of the edf ($ZA_n$ and $ZC_n$) or by the distribution function under the null hypothesis with estimated parameters ($ZB_n$).

Next we consider the Hellinger distance ($M_{m,n}$) and Kullback-Leibler divergence ($KL_{n,m}$) tests which are based on the concept of entropy. Furthermore, we review tests based on phi-divergence. These tests are based on four distance measures; the Kullback-Leibler distance ($DK_n$), the Hellinger distance ($DH_n$), the Jeffreys divergence distance ($DJ_n$) as well as the Total variation distance ($DT_n$).

Although the Pareto distribution does not have a close form expression for its characteristic function, we include a test, $S_{n,a}$, utilising the characteristic function of the uniform distribution. We also discuss a test involving the Mellin transform ($G_n$) as well as a test based on the fact that the Pareto distribution has a constant inequality curve ($TS_n$). Finally, we consider a number of tests utilising different characterisations of the Pareto distribution ($T_n$, $I_{n,2}$, $I_{n,3}$, $I_n^{[1]}$ and $I_n^{[2]}$).

For the Monte Carlo simulation, we consider eight different distributions (with various parameter settings) under the alternative hypothesis. Some of the tests utilised require parameter estimation. To this end, we consider the maximum likelihood estimators (MLE) and the adjusted method of moments estimators (MMEs). The power performance of the tests are considered in the case where only the shape parameter of the Pareto distribution requires estimation as well as in the case where both the shape and scale parameters are unknown.

The numerical powers of the various test statistics are investigated and compared using a Monte Carlo study. This study shows that $KL_{n,10}$ and $DK_n$ produces impressive power results against a range of alternative distributions when using MLE in order to estimate the parameters of the Pareto distribution. In the case where MMEs are used to perform parameter estimation, the $G_{n,2}$ test produces the highest powers followed by $MA_n$. It should, however, be note that $G_{n,2}$ produces the lowest powers against $W(0.5)$, $LN(2.5)$ and the tilted Pareto distribution. When taking all of the above into account, we recommend using $DK_n$ together with MLE when testing for the Pareto distribution in practice.

\section{Appendix}

This appendix contains the numerical results pertaining to samples of size $n=30$. The interpretation of these results are generally similar to those presented in the main text for $n=20$.

\begin{landscape}
\begin{table}[!htbp] \centering 
  \caption{Numerical powers when estimating 1 parameter using MLE with $n=30$} 
  \label{tbl:5} 
  \resizebox{\columnwidth}{!}{
\begin{tabular}{@{\extracolsep{5pt}} cccccccccccccccccccc} 
\\[-1.8ex]\hline 
\hline \\[-1.8ex]
 & $KS_n$ & $CM_n$ & $AD_n$ & $MA_n$ & $ZA_n$ & $ZB_n$ & $ZC_n$ & $KL_{n,1}$ & $KL_{n,10}$ & $DK_n$ & $S_{n,0.5}$ & $S_{n,1}$ & $G_{n,0.5}$ & $G_{n,2}$ & $T_n$ & $I_{n,2}$ & $I_{n,3}$ & $I_n^{[1]}$ & $I_n^{[2]}$ \\ 
\hline \\[-1.8ex] 
$P(1,1)$            & {5} & {5} & {5} & {5} & {5} & {5} & {5} & {5} & {5} & {5} & {5} & {5} & {5} & {5} & {5} & {5} & {5} & {5} & {5} \\
$P(2,1)$            & {5} & {5} & {5} & {5} & {5} & {5} & {5} & {5} & {5} & {5} & {5} & {5} & {5} & {5} & {5} & {5} & {5} & {5} & {5} \\
$P(5,1)$            & {5} & {5} & {5} & {5} & {5} & {5} & {5} & {5} & {5} & {5} & {5} & {5} & {5} & {5} & {5} & {5} & {5} & {5} & {5} \\
$P(10,1)$           & {5} & {5} & {5} & {5} & {5} & {5} & {5} & {5} & {5} & {5} & {5} & {5} & {5} & {5} & {5} & {5} & {5} & {5} & {5} \\
$\Gamma{(0.5)}$  & 32    & 35    & \cellcolor{black!15}\textbf{60} & 27    & 56    & \cellcolor{black!15}\textbf{63} & 2     & 25    & 1     & 1     & 36    & 32    & 57    & 32    & 37    & 45    & 55    & 59    & 9 \\
$\Gamma{(0.8)}$  & 12    & 14    & 13    & 17    & 14    & 15    & 24    & 11    & \cellcolor{black!15}\textbf{26} & \cellcolor{black!15}\textbf{30} & 14    & 14    & 8     & 12    & 14    & 10    & 8     & 7     & 20 \\
$\Gamma{(1)}$    & 35    & 44    & 39    & 47    & 42    & 39    & 52    & 21    & \cellcolor{black!15}\textbf{60} & \cellcolor{black!15}\textbf{68} & 43    & 47    & 29    & 45    & 48    & 37    & 29    & 21    & 44 \\
$\Gamma{(1.2)}$  & 64    & 76    & 73    & 77    & 76    & 74    & 76    & 40    & \cellcolor{black!15}\textbf{84} & \cellcolor{black!15}\textbf{91} & 75    & 80    & 64    & 80    & 81    & 71    & 62    & 52    & 64 \\
$W(0.5)$          & 48    & 52    & \cellcolor{black!15}\textbf{75} & 43    & 72    & \cellcolor{black!15}\textbf{77} & 2     & 39    & 1     & 0     & 53    & 49    & 74    & 48    & 54    & 61    & 71    & \cellcolor{black!15}\textbf{75} & 14 \\
$W(0.8)$          & 10    & 11    & 11    & 14    & 13    & 14    & 20    & 11    & \cellcolor{black!15}\textbf{22} & \cellcolor{black!15}\textbf{24} & 11    & 10    & 6     & 8     & 11    & 8     & 6     & 7     & 16 \\
$W(1.2)$          & 69    & 81    & 78    & 82    & 81    & 79    & 81    & 45    & \cellcolor{black!15}\textbf{88} & \cellcolor{black!15}\textbf{94} & 80    & 84    & 69    & 84    & 85    & 76    & 68    & 57    & 69 \\
$W(1.5)$          & 95    & \cellcolor{black!15}\textbf{99} & \cellcolor{black!15}\textbf{99} & \cellcolor{black!15}\textbf{99} & \cellcolor{black!15}\textbf{99} & \cellcolor{black!15}\textbf{99} & 98    & 81    & \cellcolor{black!15}\textbf{99} & \cellcolor{black!15}\textbf{100} & \cellcolor{black!15}\textbf{99} & \cellcolor{black!15}\textbf{99} & 97    & \cellcolor{black!15}\textbf{99} & \cellcolor{black!15}\textbf{99} & 98    & 96    & 93    & 89 \\
$LN(1)$           & 77    & 86    & 87    & 77    & \cellcolor{black!15}\textbf{96} & \cellcolor{black!15}\textbf{94} & 52    & 50    & 74    & 79    & 85    & 88    & \cellcolor{black!15}\textbf{94} & 91    & 90    & 86    & 87    & 84    & 28 \\
$LN(1.2)$         & 36    & 44    & 44    & 35    & \cellcolor{black!15}\textbf{64} & \cellcolor{black!15}\textbf{59} & 28    & 23    & 45    & 49    & 43    & 48    & 54    & 53    & 52    & 44    & 44    & 41    & 15 \\
$LN(1.5)$         & 8     & 9     & 8     & 8     & 15    & 12    & 10    & 9     & \cellcolor{black!15}\textbf{16} & \cellcolor{black!15}\textbf{16} & 9     & 10    & 10    & 10    & 11    & 8     & 7     & 7     & 7 \\
$LN(2.5)$         & 39    & 43    & 57    & 34    & 38    & 42    & 2     & 20    & 1     & 0     & 44    & 41    & \cellcolor{black!15}\textbf{60} & 41    & 43    & 49    & \cellcolor{black!15}\textbf{58} & \cellcolor{black!15}\textbf{58} & 16 \\
$LFR(0.2)$        & 46    & 56    & 51    & 60    & 56    & 52    & 67    & 28    & \cellcolor{black!15}\textbf{74} & \cellcolor{black!15}\textbf{80} & 54    & 59    & 38    & 58    & 60    & 48    & 37    & 28    & 57 \\
$LFR(0.5)$        & 54    & 66    & 61    & 71    & 65    & 61    & 77    & 35    & \cellcolor{black!15}\textbf{83} & \cellcolor{black!15}\textbf{89} & 64    & 69    & 46    & 67    & 70    & 58    & 45    & 36    & 67 \\
$LFR(0.8)$        & 60    & 73    & 68    & 77    & 70    & 66    & 82    & 38    & \cellcolor{black!15}\textbf{87} & \cellcolor{black!15}\textbf{92} & 71    & 75    & 51    & 73    & 76    & 65    & 52    & 41    & 73 \\
$LFR(1)$          & 63    & 75    & 71    & 80    & 72    & 69    & 83    & 41    & \cellcolor{black!15}\textbf{88} & \cellcolor{black!15}\textbf{93} & 73    & 77    & 54    & 75    & 79    & 67    & 55    & 44    & 75 \\
$BE(0.5)$         & 27    & 30    & \cellcolor{black!15}\textbf{54} & 23    & 53    & \cellcolor{black!15}\textbf{60} & 3     & 25    & 2     & 2     & 31    & 25    & 49    & 24    & 29    & 36    & 46    & 53    & 6 \\
$BE(0.8)$         & 14    & 16    & 14    & 20    & 15    & 16    & 28    & 12    & \cellcolor{black!15}\textbf{30} & \cellcolor{black!15}\textbf{34} & 16    & 16    & 8     & 13    & 16    & 11    & 8     & 6     & 23 \\
$BE(1)$           & 35    & 43    & 38    & 46    & 41    & 38    & 52    & 21    & \cellcolor{black!15}\textbf{60} & \cellcolor{black!15}\textbf{68} & 42    & 46    & 29    & 44    & 47    & 36    & 28    & 21    & 44 \\
$BE(1.5)$         & 85    & 94    & 93    & 94    & 95    & 94    & 89    & 62    & 95    & \cellcolor{black!15}\textbf{98} & 93    & \cellcolor{black!15}\textbf{96} & 90    & \cellcolor{black!15}\textbf{96} & \cellcolor{black!15}\textbf{96} & 92    & 87    & 81    & 75 \\
$TP(0.5)$         & 9     & 9     & 7     & 9     & 10    & 9     & 9     & 7     & \cellcolor{black!15}\textbf{12} & \cellcolor{black!15}\textbf{13} & 9     & 9     & 8     & 9     & 11    & 9     & 7     & 7     & 7 \\
$TP(1,1)$           & 15    & 18    & 15    & 16    & 17    & 15    & 14    & 9     & 19    & \cellcolor{black!15}\textbf{24} & 18    & 18    & 14    & 17    & \cellcolor{black!15}\textbf{21} & 16    & 13    & 12    & 13 \\
$TP(2,1)$           & 30    & 37    & 32    & 33    & 34    & 31    & 27    & 14    & 36    & \cellcolor{black!15}\textbf{44} & 36    & 38    & 29    & 37    & \cellcolor{black!15}\textbf{41} & 34    & 28    & 24    & 24 \\
$TP(3)$           & 45    & 54    & 48    & 51    & 48    & 46    & 39    & 21    & 49    & \cellcolor{black!15}\textbf{60} & 54    & 56    & 43    & 55    & \cellcolor{black!15}\textbf{58} & 50    & 42    & 37    & 35 \\
$D(0.2)$          & 16    & 19    & 16    & 16    & 22    & 19    & 14    & 10    & 21    & \cellcolor{black!15}\textbf{24} & 19    & 20    & 18    & 20    & \cellcolor{black!15}\textbf{23} & 18    & 16    & 15    & 10 \\
$D(0.4)$          & 42    & 50    & 47    & 44    & 55    & 53    & 34    & 22    & 49    & \cellcolor{black!15}\textbf{56} & 50    & 54    & 50    & \cellcolor{black!15}\textbf{56} & \cellcolor{black!15}\textbf{57} & 49    & 46    & 41    & 25 \\
$D(0.6)$          & 70    & 81    & 79    & 76    & 84    & 83    & 59    & 41    & 75    & 84    & 80    & 85    & 82    & \cellcolor{black!15}\textbf{86} & \cellcolor{black!15}\textbf{86} & 79    & 76    & 71    & 43 \\
$D(0.8)$          & 89    & 95    & 95    & 93    & \cellcolor{black!15}\textbf{97} & 96    & 79    & 62    & 91    & 96    & 95    & \cellcolor{black!15}\textbf{97} & 95    & \cellcolor{black!15}\textbf{97} & \cellcolor{black!15}\textbf{97} & 94    & 93    & 89    & 58 \\
$HN(0.8)$         & 68    & 79    & 75    & 84    & 77    & 73    & 87    & 46    & \cellcolor{black!15}\textbf{91} & \cellcolor{black!15}\textbf{95} & 78    & 81    & 57    & 78    & 82    & 72    & 59    & 47    & 79 \\
$HN(1)$           & 73    & 85    & 81    & 88    & 83    & 79    & 92    & 52    & \cellcolor{black!15}\textbf{94} & \cellcolor{black!15}\textbf{97} & 83    & 86    & 63    & 84    & 87    & 77    & 64    & 53    & 84 \\
\hline \\[-1.8ex] 
\end{tabular} 
}
\end{table}

\clearpage

\end{landscape}

\begin{landscape}
\begin{table}[!htbp] \centering 
  \caption{Numerical powers when estimating 1 parameter using MME with $n=30$} 
  \label{tbl:6} 
  \resizebox{\columnwidth}{!}{
\begin{tabular}{@{\extracolsep{5pt}} cccccccccccccccccccc} 
\\[-1.8ex]\hline 
\hline \\[-1.8ex]
 & $KS_n$ & $CM_n$ & $AD_n$ & $MA_n$ & $ZA_n$ & $ZB_n$ & $ZC_n$ & $KL_{n,1}$ & $KL_{n,10}$ & $DK_n$ & $S_{n,0.5}$ & $S_{n,1}$ & $G_{n,0.5}$ & $G_{n,2}$ & $T_n$ & $I_{n,2}$ & $I_{n,3}$ & $I_n^{[1]}$ & $I_n^{[2]}$ \\ 
\hline \\[-1.8ex] 
$P(1,1)$            & 10    & 11    & 12    & {14} & 10    & 12    & {13} & 6     & 4     & 0     & 10    & 11    & 6     & 11    & 5     & 5     & 5     & 5     & 5 \\
$P(2,1)$            & {5} & {5} & {5} & {5} & {5} & {5} & {5} & {5} & {5} & 2     & {5} & {5} & {5} & {5} & {5} & {5} & {5} & {5} & {5} \\
$P(5,1)$            & {5} & {5} & {5} & {5} & {5} & {5} & {5} & {5} & {5} & 4     & {5} & {5} & {5} & {5} & {5} & {5} & {5} & {5} & {5} \\
$P(10,1)$           & {5} & {5} & {5} & {5} & {5} & {5} & {5} & {5} & {5} & {5} & {5} & {5} & {5} & {5} & {5} & {5} & {5} & {5} & {5} \\
$\Gamma{(0.5)}$   & 21    & 21    & 48    & 15    & 52    & \cellcolor{black!15}\textbf{61} & 2     & 25    & 1     & 0     & 24    & 17    & 21    & 6     & 37    & 45    & 55    & \cellcolor{black!15}\textbf{59} & 9 \\
$\Gamma{(0.8)}$   & 19    & 22    & 20    & \cellcolor{black!15}\textbf{27} & 13    & 17    & 16    & 11    & 22    & 3     & 21    & 21    & 15    & \cellcolor{black!15}\textbf{25} & 14    & 10    & 8     & 7     & 20 \\
$\Gamma{(1)}$     & 53    & 61    & 59    & \cellcolor{black!15}\textbf{65} & 47    & 50    & 32    & 24    & 55    & 11    & 60    & 62    & 54    & \cellcolor{black!15}\textbf{67} & 48    & 37    & 29    & 21    & 44 \\
$\Gamma{(1.2)}$   & 82    & 89    & 89    & \cellcolor{black!15}\textbf{90} & 82    & 84    & 48    & 47    & 83    & 25    & 88    & \cellcolor{black!15}\textbf{90} & 86    & \cellcolor{black!15}\textbf{93} & 81    & 71    & 62    & 52    & 64 \\
$W(0.5)$          & 22    & 20    & 54    & 14    & 61    & \cellcolor{black!15}\textbf{71} & 1     & 38    & 0     & 0     & 25    & 14    & 35    & 6     & 54    & 61    & \cellcolor{black!15}\textbf{71} & \cellcolor{black!15}\textbf{75} & 14 \\
$W(0.8)$          & 20    & 21    & 21    & \cellcolor{black!15}\textbf{28} & 12    & 16    & 10    & 11    & 18    & 1     & 21    & 21    & 12    & \cellcolor{black!15}\textbf{24} & 11    & 8     & 6     & 7     & 16 \\
$W(1.2)$          & 83    & 90    & 89    & \cellcolor{black!15}\textbf{91} & 83    & 85    & 60    & 48    & 87    & 43    & 89    & \cellcolor{black!15}\textbf{91} & 89    & \cellcolor{black!15}\textbf{93} & 85    & 76    & 68    & 57    & 69 \\
$W(1.5)$          & 98    & \cellcolor{black!15}\textbf{100} & \cellcolor{black!15}\textbf{100} & \cellcolor{black!15}\textbf{100} & 99    & 99    & 91    & 85    & 99    & 92    & \cellcolor{black!15}\textbf{100} & \cellcolor{black!15}\textbf{100} & \cellcolor{black!15}\textbf{100} & \cellcolor{black!15}\textbf{100} & 99    & 98    & 96    & 93    & 89 \\
$LN(1)$           & 89    & 94    & 95    & 89    & \cellcolor{black!15}\textbf{99} & \cellcolor{black!15}\textbf{98} & 18    & 59    & 72    & 17    & 93    & 95    & 97    & 95    & 90    & 86    & 87    & 84    & 28 \\
$LN(1.2)$         & 55    & 67    & 69    & 61    & \cellcolor{black!15}\textbf{78} & \cellcolor{black!15}\textbf{75} & 6     & 29    & 39    & 4     & 63    & 68    & 71    & 72    & 52    & 44    & 44    & 41    & 15 \\
$LN(1.5)$         & 22    & 27    & 29    & \cellcolor{black!15}\textbf{31} & 29    & 28    & 2     & 13    & 12    & 1     & 25    & 28    & 21    & \cellcolor{black!15}\textbf{31} & 11    & 8     & 7     & 7     & 7 \\
$LN(2.5)$         & 13    & 10    & 31    & 31    & 31    & 45    & 39    & 27    & 0     & 0     & 14    & 7     & 10    & 4     & 43    & 49    & \cellcolor{black!15}\textbf{58} & \cellcolor{black!15}\textbf{58} & 16 \\
$LFR(0.2)$        & 61    & 70    & 69    & \cellcolor{black!15}\textbf{74} & 56    & 59    & 50    & 31    & 71    & 22    & 69    & 72    & 65    & \cellcolor{black!15}\textbf{76} & 60    & 48    & 37    & 28    & 57 \\
$LFR(0.5)$        & 69    & 77    & 76    & \cellcolor{black!15}\textbf{81} & 64    & 68    & 63    & 37    & \cellcolor{black!15}\textbf{81} & 38    & 76    & 78    & 73    & \cellcolor{black!15}\textbf{82} & 70    & 58    & 45    & 36    & 67 \\
$LFR(0.8)$        & 73    & 81    & 79    & 84    & 69    & 72    & 70    & 41    & \cellcolor{black!15}\textbf{85} & 50    & 80    & 82    & 78    & \cellcolor{black!15}\textbf{85} & 76    & 65    & 52    & 41    & 73 \\
$LFR(1)$          & 75    & 83    & 81    & \cellcolor{black!15}\textbf{86} & 72    & 75    & 74    & 42    & \cellcolor{black!15}\textbf{87} & 58    & 82    & 84    & 81    & \cellcolor{black!15}\textbf{86} & 79    & 67    & 55    & 44    & 75 \\
$BE(0.5)$         & 15    & 15    & 42    & 12    & 46    & \cellcolor{black!15}\textbf{57} & 3     & 25    & 1     & 0     & 18    & 11    & 15    & 4     & 29    & 36    & 46    & \cellcolor{black!15}\textbf{53} & 6 \\
$BE(0.8)$         & 23    & 26    & 25    & \cellcolor{black!15}\textbf{33} & 16    & 19    & 18    & 12    & 26    & 3     & 26    & 26    & 18    & \cellcolor{black!15}\textbf{30} & 16    & 11    & 8     & 6     & 23 \\
$BE(1)$           & 53    & 61    & 59    & \cellcolor{black!15}\textbf{64} & 46    & 49    & 32    & 24    & 55    & 11    & 59    & 62    & 53    & \cellcolor{black!15}\textbf{67} & 47    & 36    & 28    & 21    & 44 \\
$BE(1.5)$         & 95    & 98    & 98    & 98    & 97    & 98    & 61    & 70    & 95    & 48    & 98    & \cellcolor{black!15}\textbf{99} & 98    & \cellcolor{black!15}\textbf{99} & 96    & 92    & 87    & 81    & 75 \\
$TP(0.5)$         & 33    & 38    & \cellcolor{black!15}\textbf{43} & \cellcolor{black!15}\textbf{46} & 37    & 40    & 19    & 15    & 8     & 0     & 37    & 39    & 21    & 38    & 11    & 9     & 7     & 7     & 7 \\
$TP(1,1)$           & 59    & 66    & \cellcolor{black!15}\textbf{71} & \cellcolor{black!15}\textbf{74} & 66    & 69    & 26    & 28    & 14    & 0     & 65    & 67    & 42    & 66    & 21    & 16    & 13    & 12    & 13 \\
$TP(2,1)$           & 88    & 93    & \cellcolor{black!15}\textbf{95} & \cellcolor{black!15}\textbf{96} & 92    & 94    & 40    & 59    & 27    & 0     & 92    & 93    & 73    & 92    & 41    & 34    & 28    & 24    & 24 \\
$TP(3)$           & 97    & \cellcolor{black!15}\textbf{99} & \cellcolor{black!15}\textbf{99} & \cellcolor{black!15}\textbf{99} & \cellcolor{black!15}\textbf{99} & \cellcolor{black!15}\textbf{99} & 55    & 81    & 40    & 0     & 98    & \cellcolor{black!15}\textbf{99} & 88    & \cellcolor{black!15}\textbf{99} & 58    & 50    & 42    & 37    & 35 \\
$D(0.2)$          & 33    & 40    & 40    & 40    & 38    & 36    & 4     & 14    & 16    & 1     & 38    & \cellcolor{black!15}\textbf{41} & 32    & \cellcolor{black!15}\textbf{43} & 23    & 18    & 16    & 15    & 10 \\
$D(0.4)$          & 67    & 75    & 76    & 72    & 74    & 73    & 6     & 31    & 42    & 4     & 74    & \cellcolor{black!15}\textbf{77} & 72    & \cellcolor{black!15}\textbf{79} & 57    & 49    & 46    & 41    & 25 \\
$D(0.6)$          & 89    & 94    & 94    & 92    & 93    & 93    & 16    & 53    & 70    & 11    & 93    & \cellcolor{black!15}\textbf{95} & 93    & \cellcolor{black!15}\textbf{95} & 86    & 79    & 76    & 71    & 43 \\
$D(0.8)$          & 97    & \cellcolor{black!15}\textbf{99} & \cellcolor{black!15}\textbf{99} & \cellcolor{black!15}\textbf{99} & \cellcolor{black!15}\textbf{99} & \cellcolor{black!15}\textbf{99} & 30    & 74    & 89    & 25    & \cellcolor{black!15}\textbf{99} & \cellcolor{black!15}\textbf{99} & \cellcolor{black!15}\textbf{99} & \cellcolor{black!15}\textbf{99} & 97    & 94    & 93    & 89    & 58 \\
$HN(0.8)$         & 77    & 85    & 84    & \cellcolor{black!15}\textbf{88} & 75    & 78    & 79    & 46    & \cellcolor{black!15}\textbf{90} & 64    & 84    & 86    & 83    & \cellcolor{black!15}\textbf{88} & 82    & 72    & 59    & 47    & 79 \\
$HN(1)$           & 85    & 91    & 90    & \cellcolor{black!15}\textbf{93} & 82    & 85    & 82    & 55    & \cellcolor{black!15}\textbf{93} & 62    & 90    & 92    & 87    & \cellcolor{black!15}\textbf{93} & 87    & 77    & 64    & 53    & 84 \\
\hline \\[-1.8ex] 
\end{tabular} 
}
\end{table}

\clearpage

\end{landscape}

\begin{landscape}
\begin{table}[!htbp] \centering 
  \caption{Numerical powers when estimating 2 parameter using MLE with $n=30$} 
  \label{tbl:7} 
  \resizebox{\columnwidth}{!}{
\begin{tabular}{@{\extracolsep{5pt}} cccccccccccccccccccc} 
\\[-1.8ex]\hline 
\hline \\[-1.8ex]
 & $KS_n$ & $CM_n$ & $AD_n$ & $MA_n$ & $ZA_n$ & $ZB_n$ & $ZC_n$ & $KL_{n,1}$ & $KL_{n,10}$ & $DK_n$ & $S_{n,0.5}$ & $S_{n,1}$ & $G_{n,0.5}$ & $G_{n,2}$ & $T_n$ & $I_{n,2}$ & $I_{n,3}$ & $I_n^{[1]}$ & $I_n^{[2]}$ \\ 
\hline \\[-1.8ex] 
$P(1,1)$            & {5} & {5} & {5} & {5} & {5} & {5} & {5} & {5} & {5} & {5} & {5} & {5} & {5} & {5} & {5} & 4     & 4     & {5} & {5} \\
$P(2,1)$            & {5} & {5} & {5} & {5} & {5} & 4     & {5} & {5} & {5} & {5} & {5} & {5} & {5} & {5} & {5} & {5} & {5} & {5} & {5} \\
$P(5,1)$            & {5} & {5} & {5} & {5} & {5} & {5} & {5} & {5} & {5} & {5} & {5} & {5} & {5} & {5} & {5} & {5} & {5} & {5} & {5} \\
$P(10,1)$           & {5} & {5} & {5} & {5} & {5} & {5} & {5} & {5} & {5} & {5} & {5} & {5} & {5} & {5} & {5} & {5} & {5} & {5} & {5} \\
$\Gamma{(0.5)}$   & 33    & 36    & \cellcolor{black!15}\textbf{55} & 27    & \cellcolor{black!15}\textbf{53} & 52    & 2     & 34    & 2     & 1     & 37    & 33    & \cellcolor{black!15}\textbf{53} & 32    & 38    & 40    & 48    & 45    & 11 \\
$\Gamma{(0.8)}$   & 10    & 11    & 7     & 14    & 11    & 7     & \cellcolor{black!15}\textbf{30} & 12    & 29    & \cellcolor{black!15}\textbf{32} & 11    & 8     & 2     & 5     & 8     & 5     & 3     & 7     & 20 \\
$\Gamma{(1)}$     & 26    & 30    & 16    & 35    & 23    & 6     & \cellcolor{black!15}\textbf{61} & 17    & 57    & \cellcolor{black!15}\textbf{66} & 29    & 28    & 3     & 21    & 28    & 17    & 8     & 21    & 43 \\
$\Gamma{(1.2)}$   & 45    & 54    & 34    & 58    & 44    & 13    & \cellcolor{black!15}\textbf{82} & 26    & 76    & \cellcolor{black!15}\textbf{86} & 52    & 53    & 9     & 44    & 53    & 37    & 22    & 40    & 63 \\
$W(0.5)$          & 49    & 53    & \cellcolor{black!15}\textbf{72} & 43    & \cellcolor{black!15}\textbf{70} & 68    & 2     & 49    & 2     & 1     & 54    & 49    & \cellcolor{black!15}\textbf{70} & 47    & 54    & 56    & 64    & 62    & 17 \\
$W(0.8)$          & 9     & 9     & 7     & 12    & 10    & 8     & \cellcolor{black!15}\textbf{25} & 12    & \cellcolor{black!15}\textbf{25} & \cellcolor{black!15}\textbf{27} & 9     & 7     & 3     & 4     & 7     & 4     & 3     & 6     & 17 \\
$W(1.2)$          & 49    & 59    & 40    & 64    & 50    & 16    & \cellcolor{black!15}\textbf{86} & 29    & 80    & \cellcolor{black!15}\textbf{88} & 58    & 59    & 11    & 50    & 59    & 42    & 26    & 45    & 68 \\
$W(1.5)$          & 77    & 85    & 72    & 88    & 79    & 43    & \cellcolor{black!15}\textbf{97} & 52    & 95    & \cellcolor{black!15}\textbf{98} & 85    & 85    & 33    & 79    & 86    & 73    & 55    & 74    & 88 \\
$LN(1)$           & 26    & 30    & 16    & 28    & 20    & 5     & \cellcolor{black!15}\textbf{46} & 12    & 40    & \cellcolor{black!15}\textbf{50} & 30    & 30    & 5     & 25    & 31    & 21    & 14    & 34    & 25 \\
$LN(1.2)$         & 12    & 13    & 6     & 12    & 8     & 2     & \cellcolor{black!15}\textbf{28} & 7     & 26    & \cellcolor{black!15}\textbf{31} & 13    & 13    & 1     & 9     & 13    & 8     & 5     & 17    & 14 \\
$LN(1.5)$         & 5     & 4     & 2     & 5     & 3     & 3     & \cellcolor{black!15}\textbf{11} & 5     & \cellcolor{black!15}\textbf{11} & \cellcolor{black!15}\textbf{13} & 4     & 4     & 2     & 3     & 4     & 3     & 2     & 5     & 7 \\
$LN(2.5)$         & 42    & 45    & \cellcolor{black!15}\textbf{58} & 36    & 51    & 45    & 1     & 27    & 1     & 1     & 47    & 44    & \cellcolor{black!15}\textbf{60} & 43    & 48    & 49    & 56    & 48    & 18 \\
$LFR(0.2)$        & 34    & 41    & 24    & 47    & 34    & 9     & \cellcolor{black!15}\textbf{74} & 22    & 70    & \cellcolor{black!15}\textbf{77} & 40    & 39    & 4     & 30    & 39    & 25    & 13    & 27    & 55 \\
$LFR(0.5)$        & 42    & 51    & 31    & 58    & 43    & 13    & \cellcolor{black!15}\textbf{83} & 27    & 79    & \cellcolor{black!15}\textbf{86} & 49    & 49    & 6     & 38    & 49    & 32    & 17    & 33    & 66 \\
$LFR(0.8)$        & 47    & 56    & 37    & 64    & 49    & 15    & \cellcolor{black!15}\textbf{87} & 30    & 83    & \cellcolor{black!15}\textbf{89} & 55    & 55    & 8     & 44    & 55    & 38    & 20    & 38    & 72 \\
$LFR(1)$          & 50    & 60    & 40    & 67    & 52    & 17    & \cellcolor{black!15}\textbf{88} & 33    & 84    & \cellcolor{black!15}\textbf{91} & 58    & 58    & 9     & 47    & 58    & 40    & 22    & 40    & 74 \\
$BE(0.5)$         & 27    & 30    & \cellcolor{black!15}\textbf{49} & 23    & \cellcolor{black!15}\textbf{49} & \cellcolor{black!15}\textbf{49} & 3     & 33    & 4     & 3     & 31    & 26    & 44    & 24    & 30    & 32    & 40    & 39    & 8 \\
$BE(0.8)$         & 12    & 12    & 7     & 16    & 12    & 7     & \cellcolor{black!15}\textbf{34} & 13    & 33    & \cellcolor{black!15}\textbf{37} & 12    & 10    & 2     & 6     & 9     & 5     & 3     & 7     & 23 \\
$BE(1)$           & 25    & 29    & 15    & 34    & 22    & 6     & \cellcolor{black!15}\textbf{61} & 17    & 56    & \cellcolor{black!15}\textbf{64} & 28    & 28    & 2     & 20    & 27    & 16    & 8     & 20    & 43 \\
$BE(1.5)$         & 60    & 70    & 52    & 73    & 61    & 24    & \cellcolor{black!15}\textbf{90} & 35    & 85    & \cellcolor{black!15}\textbf{92} & 69    & 71    & 19    & 63    & 71    & 55    & 38    & 59    & 73 \\
$TP(0.5)$         & 6     & 5     & 3     & 6     & 4     & 3     & \cellcolor{black!15}\textbf{12} & 6     & 11    & \cellcolor{black!15}\textbf{13} & 5     & 5     & 1     & 3     & 5     & 3     & 2     & 6     & 7 \\
$TP(1,1)$           & 10    & 10    & 4     & 10    & 6     & 3     & \cellcolor{black!15}\textbf{19} & 7     & 17    & \cellcolor{black!15}\textbf{22} & 10    & 9     & 1     & 6     & 10    & 6     & 3     & 11    & 12 \\
$TP(2,1)$           & 19    & 21    & 10    & 22    & 13    & 4     & \cellcolor{black!15}\textbf{36} & 10    & 31    & \cellcolor{black!15}\textbf{39} & 21    & 20    & 2     & 15    & 21    & 13    & 7     & 21    & 23 \\
$TP(3)$           & 31    & 35    & 19    & 34    & 21    & 6     & \cellcolor{black!15}\textbf{51} & 14    & 42    & \cellcolor{black!15}\textbf{54} & 34    & 33    & 4     & 26    & 34    & 24    & 14    & 32    & 35 \\
$D(0.2)$          & 8     & 9     & 4     & 8     & 5     & 3     & \cellcolor{black!15}\textbf{18} & 6     & 16    & \cellcolor{black!15}\textbf{20} & 8     & 8     & 1     & 5     & 9     & 5     & 3     & 12    & 10 \\
$D(0.4)$          & 21    & 24    & 11    & 23    & 15    & 4     & \cellcolor{black!15}\textbf{41} & 10    & 35    & \cellcolor{black!15}\textbf{44} & 24    & 24    & 3     & 18    & 25    & 16    & 10    & 26    & 24 \\
$D(0.6)$          & 39    & 45    & 27    & 44    & 31    & 9     & \cellcolor{black!15}\textbf{64} & 17    & 55    & \cellcolor{black!15}\textbf{67} & 45    & 45    & 9     & 38    & 46    & 34    & 22    & 44    & 41 \\
$D(0.8)$          & 54    & 63    & 44    & 62    & 49    & 17    & \cellcolor{black!15}\textbf{79} & 25    & 70    & \cellcolor{black!15}\textbf{82} & 62    & 63    & 18    & 57    & 63    & 50    & 36    & 59    & 56 \\
$HN(0.8)$         & 55    & 65    & 46    & 72    & 58    & 20    & \cellcolor{black!15}\textbf{91} & 37    & 88    & \cellcolor{black!15}\textbf{93} & 63    & 63    & 10    & 52    & 63    & 45    & 25    & 44    & 78 \\
$HN(1)$           & 60    & 71    & 52    & 78    & 65    & 25    & \cellcolor{black!15}\textbf{94} & 42    & 92    & \cellcolor{black!15}\textbf{95} & 69    & 69    & 14    & 58    & 70    & 52    & 30    & 49    & 83 \\
\hline \\[-1.8ex] 
\end{tabular} 
}
\end{table}

\clearpage

\end{landscape}

\begin{landscape}
\begin{table}[!htbp] \centering 
  \caption{Numerical powers when estimating 2 parameter using MME with $n=30$} 
  \label{tbl:8} 
  \resizebox{\columnwidth}{!}{
\begin{tabular}{@{\extracolsep{5pt}} cccccccccccccccccccc} 
\\[-1.8ex]\hline 
\hline \\[-1.8ex]
 & $KS_n$ & $CM_n$ & $AD_n$ & $MA_n$ & $ZA_n$ & $ZB_n$ & $ZC_n$ & $KL_{n,1}$ & $KL_{n,10}$ & $DK_n$ & $S_{n,0.5}$ & $S_{n,1}$ & $G_{n,0.5}$ & $G_{n,2}$ & $T_n$ & $I_{n,2}$ & $I_{n,3}$ & $I_n^{[1]}$ & $I_n^{[2]}$ \\ 
\hline \\[-1.8ex] 
$P(1,1)$            & 10    & 11    & 12    & {14} & 13    & {14} & 13    & 6     & 4     & 0     & 10    & 11    & 6     & 11    & 5     & 4     & 4     & 5     & 5 \\
$P(2,1)$            & {5} & {5} & {5} & {5} & {5} & {5} & {5} & {5} & {5} & 2     & {5} & {5} & {5} & {5} & {5} & {5} & {5} & {5} & {5} \\
$P(5,1)$            & {5} & {5} & {5} & {5} & {5} & {5} & {5} & {5} & {5} & 4     & {5} & {5} & {5} & {5} & {5} & {5} & {5} & {5} & {5} \\
$P(10,1)$           & {5} & {5} & {5} & {5} & {5} & {5} & {5} & {5} & {5} & {5} & {5} & {5} & {5} & {5} & {5} & {5} & {5} & {5} & {5} \\
$\Gamma{(0.5)}$   & 15    & 15    & 25    & 11    & 18    & 21    & 2     & 34    & 2     & 0     & 16    & 11    & 14    & 4     & 38    & 40    & \cellcolor{black!15}\textbf{48} & \cellcolor{black!15}\textbf{45} & 11 \\
$\Gamma{(0.8)}$   & 21    & 23    & 23    & \cellcolor{black!15}\textbf{28} & 19    & 23    & 17    & 12    & 25    & 3     & 23    & 24    & 18    & \cellcolor{black!15}\textbf{28} & 8     & 5     & 3     & 7     & 20 \\
$\Gamma{(1)}$     & 50    & 58    & 57    & \cellcolor{black!15}\textbf{61} & 53    & 58    & 30    & 18    & 51    & 9     & 56    & 59    & 52    & \cellcolor{black!15}\textbf{64} & 28    & 17    & 8     & 21    & 43 \\
$\Gamma{(1.2)}$   & 73    & 81    & 82    & \cellcolor{black!15}\textbf{83} & 79    & 82    & 43    & 29    & 73    & 20    & 80    & \cellcolor{black!15}\textbf{83} & 78    & \cellcolor{black!15}\textbf{86} & 53    & 37    & 22    & 40    & 63 \\
$W(0.5)$          & 16    & 14    & 28    & 12    & 25    & 30    & 1     & 48    & 1     & 0     & 18    & 9     & 21    & 4     & 54    & 56    & \cellcolor{black!15}\textbf{64} & \cellcolor{black!15}\textbf{62} & 17 \\
$W(0.8)$          & 21    & 24    & 24    & \cellcolor{black!15}\textbf{30} & 19    & 23    & 11    & 12    & 21    & 1     & 23    & 24    & 16    & \cellcolor{black!15}\textbf{28} & 7     & 4     & 3     & 6     & 17 \\
$W(1.2)$          & 74    & 82    & 81    & \cellcolor{black!15}\textbf{83} & 80    & \cellcolor{black!15}\textbf{83} & 54    & 30    & 77    & 33    & 80    & \cellcolor{black!15}\textbf{83} & 80    & \cellcolor{black!15}\textbf{86} & 59    & 42    & 26    & 45    & 68 \\
$W(1.5)$          & 90    & 95    & 95    & 95    & 95    & \cellcolor{black!15}\textbf{96} & 81    & 53    & 95    & 70    & 94    & 95    & 95    & \cellcolor{black!15}\textbf{96} & 86    & 73    & 55    & 74    & 88 \\
$LN(1)$           & 50    & 58    & 58    & 54    & 57    & 58    & 12    & 13    & 35    & 9     & 56    & \cellcolor{black!15}\textbf{59} & 58    & \cellcolor{black!15}\textbf{62} & 31    & 21    & 14    & 34    & 25 \\
$LN(1.2)$         & 32    & 39    & 40    & 39    & 39    & 39    & 5     & 9     & 20    & 3     & 37    & \cellcolor{black!15}\textbf{41} & 36    & \cellcolor{black!15}\textbf{44} & 13    & 8     & 5     & 17    & 14 \\
$LN(1.5)$         & 17    & 20    & 22    & \cellcolor{black!15}\textbf{25} & 20    & 21    & 2     & 7     & 8     & 1     & 19    & 21    & 13    & \cellcolor{black!15}\textbf{23} & 4     & 3     & 2     & 5     & 7 \\
$LN(2.5)$         & 14    & 11    & 27    & 33    & 38    & 46    & 39    & 36    & 0     & 0     & 14    & 8     & 5     & 6     & 48    & \cellcolor{black!15}\textbf{49} & \cellcolor{black!15}\textbf{56} & 48    & 18 \\
$LFR(0.2)$        & 58    & 67    & 66    & \cellcolor{black!15}\textbf{71} & 64    & 69    & 47    & 24    & 67    & 19    & 65    & 68    & 63    & \cellcolor{black!15}\textbf{73} & 39    & 25    & 13    & 27    & 55 \\
$LFR(0.5)$        & 65    & 74    & 73    & \cellcolor{black!15}\textbf{77} & 72    & \cellcolor{black!15}\textbf{77} & 60    & 29    & \cellcolor{black!15}\textbf{77} & 32    & 72    & 75    & 70    & \cellcolor{black!15}\textbf{79} & 49    & 32    & 17    & 33    & 66 \\
$LFR(0.8)$        & 69    & 77    & 77    & 80    & 77    & \cellcolor{black!15}\textbf{81} & 66    & 31    & \cellcolor{black!15}\textbf{81} & 43    & 76    & 78    & 75    & \cellcolor{black!15}\textbf{82} & 55    & 38    & 20    & 38    & 72 \\
$LFR(1)$          & 70    & 79    & 79    & 82    & 79    & \cellcolor{black!15}\textbf{83} & 70    & 33    & \cellcolor{black!15}\textbf{83} & 49    & 78    & 80    & 78    & \cellcolor{black!15}\textbf{83} & 58    & 40    & 22    & 40    & 74 \\
$BE(0.5)$         & 11    & 10    & 19    & 10    & 14    & 18    & 3     & 33    & 3     & 0     & 12    & 7     & 9     & 3     & 30    & 32    & \cellcolor{black!15}\textbf{40} & \cellcolor{black!15}\textbf{39} & 8 \\
$BE(0.8)$         & 24    & 28    & 27    & \cellcolor{black!15}\textbf{34} & 23    & 28    & 18    & 13    & 29    & 3     & 27    & 28    & 22    & \cellcolor{black!15}\textbf{33} & 9     & 5     & 3     & 7     & 23 \\
$BE(1)$           & 49    & 57    & 57    & \cellcolor{black!15}\textbf{61} & 53    & 57    & 30    & 18    & 51    & 9     & 56    & 59    & 52    & \cellcolor{black!15}\textbf{63} & 27    & 16    & 8     & 20    & 43 \\
$BE(1.5)$         & 84    & 90    & 90    & 90    & 88    & 90    & 52    & 39    & 83    & 32    & 89    & \cellcolor{black!15}\textbf{91} & 89    & \cellcolor{black!15}\textbf{92} & 71    & 55    & 38    & 59    & 73 \\
$TP(0.5)$         & 31    & 36    & 41    & \cellcolor{black!15}\textbf{44} & 43    & \cellcolor{black!15}\textbf{44} & 19    & 12    & 7     & 0     & 35    & 37    & 20    & 36    & 5     & 3     & 2     & 6     & 7 \\
$TP(1,1)$           & 55    & 62    & 68    & 70    & \cellcolor{black!15}\textbf{71} & \cellcolor{black!15}\textbf{71} & 25    & 22    & 12    & 0     & 61    & 63    & 38    & 62    & 10    & 6     & 3     & 11    & 12 \\
$TP(2,1)$           & 84    & 89    & 92    & \cellcolor{black!15}\textbf{93} & \cellcolor{black!15}\textbf{93} & \cellcolor{black!15}\textbf{93} & 37    & 47    & 22    & 0     & 88    & 89    & 67    & 89    & 21    & 13    & 7     & 21    & 23 \\
$TP(3)$           & 95    & 97    & 98    & 98    & \cellcolor{black!15}\textbf{99} & \cellcolor{black!15}\textbf{99} & 50    & 68    & 31    & 0     & 96    & 97    & 83    & 97    & 34    & 24    & 14    & 32    & 35 \\
$D(0.2)$          & 28    & 33    & \cellcolor{black!15}\textbf{34} & 33    & 32    & 32    & 4     & 8     & 12    & 1     & 31    & \cellcolor{black!15}\textbf{34} & 25    & \cellcolor{black!15}\textbf{35} & 9     & 5     & 3     & 12    & 10 \\
$D(0.4)$          & 51    & 60    & 60    & 58    & 57    & 57    & 6     & 14    & 29    & 3     & 58    & \cellcolor{black!15}\textbf{61} & 53    & \cellcolor{black!15}\textbf{63} & 25    & 16    & 10    & 26    & 24 \\
$D(0.6)$          & 70    & 77    & 78    & 76    & 75    & 75    & 13    & 22    & 48    & 8     & 76    & \cellcolor{black!15}\textbf{79} & 73    & \cellcolor{black!15}\textbf{81} & 46    & 34    & 22    & 44    & 41 \\
$D(0.8)$          & 81    & 87    & 87    & 86    & 85    & 86    & 25    & 32    & 66    & 17    & 87    & \cellcolor{black!15}\textbf{88} & 85    & \cellcolor{black!15}\textbf{90} & 63    & 50    & 36    & 59    & 56 \\
$HN(0.8)$         & 73    & 81    & 81    & \cellcolor{black!15}\textbf{85} & 82    & \cellcolor{black!15}\textbf{85} & 75    & 35    & \cellcolor{black!15}\textbf{87} & 55    & 80    & 82    & 80    & \cellcolor{black!15}\textbf{85} & 63    & 45    & 25    & 44    & 78 \\
$HN(1)$           & 81    & 88    & 88    & 90    & 87    & 90    & 78    & 43    & \cellcolor{black!15}\textbf{91} & 52    & 87    & 88    & 85    & \cellcolor{black!15}\textbf{91} & 70    & 52    & 30    & 49    & 83 \\
\hline \\[-1.8ex] 
\end{tabular} 
}
\end{table}

\clearpage

\end{landscape}

\bibliographystyle{apa-good}
\bibliography{References}

\end{document}